\documentclass[aps,pre,twocolumn,amsmath,amsfonts,showpacs]{revtex4}
\newcommand{\bec}[1]{\mbox{\boldmath $ #1$}}
\usepackage{graphicx}
\begin{document}
\title{Turbulent Diffusion and Turbulent Thermal Diffusion of Aerosols in Stratified Atmospheric Flows}
\author{M. Sofiev}
\email{mikhail.sofiev@fmi.fi}
\author{V. Sofieva}
\email{Viktoria.Sofieva@fmi.fi}
\affiliation{Finnish Meteorological Institute, PL 503 (Erik Palmenin aukio 1), 00101 Helsinki, Finland}
\author{T. Elperin}
\email{elperin@bgu.ac.il}
\author{N. Kleeorin}
\email{nat@bgu.ac.il}
\author{I. Rogachevskii}
\email{gary@bgu.ac.il}
\affiliation{The Pearlstone Center for Aeronautical Engineering
Studies, Department of Mechanical Engineering,
Ben-Gurion University of the Negev, P.O.Box 653, Beer-Sheva 84105,  Israel}
\author{S. S. Zilitinkevich}
\email{Sergej.Zilitinkevich@fmi.fi}
\affiliation{Finnish Meteorological Institute and Division of Atmospheric Sciences, University of Helsinki, PL 503 (Erik Palmenin aukio 1), 00101 Helsinki, Finland; Nansen Environmental and Remote Sensing Centre / Bjerknes Centre for Climate Research, Bergen, Norway}
\date{\today}
\begin{abstract}
The paper analyzes the phenomenon of turbulent thermal diffusion in the Earth atmosphere, its relation to the turbulent diffusion and its potential impact on aerosol distribution. This phenomenon was predicted theoretically more than 10 years ago and detected recently in the laboratory experiments. This effect causes a non-diffusive flux of aerosols in the direction of the heat flux and results in formation of long-living aerosol layers in the vicinity of temperature inversions. We applied the theory of turbulent thermal diffusion to the GOMOS aerosol observations near the tropopause in order to explain the shape of aerosol vertical profiles with elevated concentrations located almost symmetrically with respect to temperature profile. We demonstrate that this theory is in good agreement with the observed profiles of aerosol concentration and temperature in the vicinity of the tropopause. In combination with the derived expression for the dependence of the turbulent thermal diffusion ratio on the turbulent diffusion, these measurements yield an independent method for determining the coefficient of turbulent diffusion at the tropopause. We also derived a practically applicable formulation for dispersion of atmospheric trace species which takes into account the phenomenon of turbulent thermal diffusion. We evaluated the impact of turbulent thermal diffusion to the lower-troposphere vertical profiles of aerosol concentration by means of  numerical dispersion modelling, and found a regular upward forcing of aerosols with coarse particles affected stronger than fine aerosols.
\end{abstract}

\maketitle

\section{Introduction}

Various aspects of turbulent diffusion of aerosols in the atmospheric flows have been extensively investigated in the past [e.g., Monin and Yaglom, 1975; Csanady, 1980; Maxey, 1987; Flagan and Seinfeld, 1988; Wyngaard, 1992; Fessler et al., 1994; Blackadar, 1997]. In particular, the turbulent diffusion (eddy diffusivity) has been comprehensively studied for the low-order closures of the turbulent dispersion equation. However, certain important features of turbulent transport of aerosols in stratified flows have been found only recently. In particular, a new phenomenon of turbulent thermal diffusion has been predicted theoretically by Elperin et al. [1996, 1997a] and detected in the laboratory experiments in stably and unstably stratified turbulent flows by Eidelman et al. [2004, 2006a, 2006b] and Buchholtz et al. [2004]. The phenomenon of turbulent thermal diffusion (TTD) in turbulent stratified flows results in the non-diffusive flux of aerosols and gaseous admixtures in the direction of the heat flux. Particles are accumulated in the vicinity of the minimum of the mean temperature of the surrounding fluid. This phenomenon causes formation of large-scale inhomogeneities in spatial distribution of aerosol particles in the vicinity of temperature inversions.

The effect of turbulent thermal diffusion has been detected in two experimental set-ups: oscillating-grids turbulence generator [Eidelman et al., 2004, 2006a; Buchholtz et al., 2004] and multi-fan turbulence generator  [Eidelman et al., 2006b]. The experiments have been performed for stably and unstably stratified fluid. In these experiments, even with strongly inhomogeneous temperature fields, particles in turbulent fluid accumulate in the regions of temperature minima, in a very good agreement with the theory of turbulent thermal diffusion.

In spite of the previous comprehensive theoretical and laboratory studies of the phenomenon of turbulent thermal diffusion, the observational evidence and quantitative evaluation of its importance in the Earth atmosphere have not been investigated until now. In the present study, we analyzed the GOMOS observations in the vicinity of the tropopause and explained the shape of aerosol vertical profiles with elevated concentrations near the minimum of temperature. The contribution of the effect of turbulent thermal diffusion to the lower-troposphere vertical profiles of the aerosol concentration was investigated via aerosol dispersion modelling.

The existing theory of turbulent thermal diffusion [Elperin et al., 1996, 1997a, 1997b, 1998, 2000a, 2000b, 2000c, 2001] does not take into account the structure of the atmospheric stratified flows. In this paper, the theoretical approach of Zilitinkevich et al. [2007, 2008] is further developed and applied to study the effect of stratification on turbulent transport of aerosols. To this end we derive the budget equation for the turbulent flux of particles in stably stratified flow. This allows determining the dependence of the turbulent diffusion and turbulent thermal diffusion coefficients on the flux Richardson number. We demonstrate that the coefficients of turbulent thermal diffusion and turbulent diffusion decrease with the increase of the flux Richardson number.

\section{Turbulent thermal diffusion and turbulent flux of aerosols}

\subsection{Mechanism of turbulent thermal diffusion}

Let us discuss the physics of the phenomenon of turbulent thermal diffusion. We consider inertial particles (aerosols) suspended in the turbulent fluid flow with large Reynolds numbers. Particle concentration   $n_p = N + n$ is characterized by the mean value, $N$, and fluctuations, $n$ (measured in m$^{-3}$). Evolution of the number density $n_p(t, {\bf r})$ of small inertial particles in a turbulent flow is determined by the following equation:
\begin{eqnarray}
{\partial n_p \over \partial t} + \bec\nabla {\bf \cdot} \, (n_p {\bf
v}) = D \,\bec{\nabla}^2 n_p \;,
\label{BB1}
\end{eqnarray}
where ${\bf v}$  is a random velocity field of the particles which they acquire in a turbulent fluid velocity field  ${\bf u}$, and $D$  is the coefficient of molecular (Brownian) diffusion. We assume here for simplicity that the mean velocity is zero, and we do not take into account the effect of particles upon the carrying fluid flow. The velocity of particles ${\bf v}$  depends on the velocity of the surrounding fluid and it can be determined from the equation of motion for a particle.  When $\rho_p \gg \rho$, this equation represents a balance of particle inertia with the fluid drag force produced by the motion of the particle relative to the surrounding fluid, $d{\bf v}/dt = ({\bf u} - {\bf v}) / \tau_s$, where $\tau_s$ is the particle Stokes time, $\rho$ is the fluid density and $\rho_p$ is the material density of a particle. Solution of the equation of motion for small particles yields:
\begin{eqnarray}
{\bf v} = {\bf u} - \tau_s \biggl[{\partial {\bf u}
\over \partial t} + ({\bf u} {\bf \cdot} \bec{\nabla}) {\bf u}
\biggr] + {\rm O}(\tau_s^2) \;,
\label{BB2}
\end{eqnarray}
[see, e.g., Maxey, 1987]. The second term in Eq.~(\ref{BB2}) describes the difference between the local fluid velocity and particle velocity arising due to the small but finite inertia of the particle. In this study we consider low Mach numbers turbulent flow with $\bec\nabla {\bf \cdot} \, {\bf u} = - \rho^{-1} \, ({\bf u} {\bf \cdot} \bec{\nabla}) \rho \not= 0$. Equation~(\ref{BB2}) for the velocity of particles and Navier-Stokes equation for the fluid for large Reynolds numbers yield
\begin{eqnarray}
\bec\nabla {\bf \cdot} \, {\bf v} &=& \bec\nabla {\bf \cdot} \, {\bf u}
- \tau_s \, \bec\nabla {\bf \cdot} \,  \biggl( {d{\bf u} \over dt} \biggr)
+ {\rm O}(\tau_s^2)
\nonumber \\
&=& - {1 \over \rho} \, ({\bf u} {\bf \cdot} \bec{\nabla}) \rho + {\tau_s \over \rho}  \,\bec{\nabla}^2 p  + {\rm O}(\tau_s^2) \;,
\label{BB3}
\end{eqnarray}
where $p$ is the fluid pressure.

The physical mechanism of the phenomenon of turbulent thermal diffusion for inertial particles can be explained as follows. Due to inertia, particles inside the turbulent eddies drift out to the boundary regions between the eddies (the regions with the decreased velocity of the turbulent fluid flow). Neglecting non-stationarity and molecular viscosity, the estimate based on the Bernoulli's law implies that these are the regions with the increased pressure of the surrounding fluid. Consequently, particles are accumulated in the regions with the maximum pressure of the turbulent fluid. Indeed, due to the inertia effect $\bec\nabla {\bf \cdot} \, {\bf v} \propto (\tau_s /\rho) \,\bec{\nabla}^2 p \not=0$  even for incompressible fluid flow [see Eq.~(\ref{BB3})]. On the other hand, for large Peclet numbers, when we can neglect the molecular diffusion of particles in Eq.~(\ref{BB1}),  $\bec\nabla {\bf \cdot} \, {\bf v} \propto - d n_p / d t$. This implies that in regions with maximum pressure of turbulent fluid (i.e., where $\bec{\nabla}^2 p < 0)$ there is accumulation of inertial particles (i.e.,  $dn_p / dt \propto - (\tau_s /\rho) \,\bec{\nabla}^2 p > 0)$. Similarly, there is an outflow of inertial particles from regions with minimum pressure of fluid. Note that in cloud physics, the effect of local accumulation of particles between turbulent eddies (in the regions with maximum fluid pressure) has been invoked extensively in order to elucidate the mechanism of rain formation [see, e.g., Pinsky and Khain, 1997; Shaw, 2003; Collins and Keswani, 2004; Wang et al., 2006; Khain et al., 2007].

In case of homogeneous and isotropic turbulence without external large-scale gradients of temperature, a drift from regions with increased (decreased) concentration of particles by a turbulent flow of fluid is equiprobable in all directions. Therefore pressure (temperature) of the fluid is not correlated with the turbulent velocity field and there exists only turbulent diffusion of particles.

Situation drastically changes in a turbulent fluid with a mean temperature gradient. In this case, the heat flux $\langle {\bf u} \, \theta \rangle$ is not zero, i.e., fluctuations of fluid temperature, $\theta$, and velocity are correlated. We consider low-Mach-number flows $(M= u/c_s \ll 1, \, c_s$ is the sound speed) and study mean-field effects. For low-Mach-number flows, the mean fluid mass flux $\langle {\bf u} \, \rho' \rangle$ is very small $(\sim O(M^2))$ [see, e.g., Chassaing et al., 2002], i.e., the fluctuations of the fluid density $\rho'$ and velocity ${\bf u}$ are weakly correlated. On the other hand, fluctuations of pressure must be correlated with the fluctuations of velocity due to a non-zero turbulent heat flux, $\langle {\bf u} \, \theta \rangle \not = 0$. Indeed, using the equation of state for an ideal gas we find that
\begin{eqnarray}
{p \over P} = {\rho' \over \rho} + {\theta \over T} \;,
\label{DS7}
\end{eqnarray}
(see below) and  $\langle {\bf u} \, p \rangle /P = \langle {\bf u} \, \theta \rangle / T$, where $P$, $\,T$ and $\rho$ are the mean fluid pressure, temperature and density, respectively. Therefore, the fluctuations of temperature and pressure are correlated and the pressure fluctuations cause fluctuations of the number density of particles. Indeed, increase (decrease) of the pressure of surrounding fluid is accompanied by accumulation (outflow) of the particles, respectively. The direction of the mean flux of particles coincides with the direction of the heat flux of temperature - towards the minimum of the mean temperature. Therefore, the particles are accumulated in this region [for more details, see Elperin et al., 1996, 1997a].

Equation~(\ref{DS7}) is obtained as follows. Averaging the ideal gas equation yields equation for the mean pressure: $P = (\kappa_b / m_\mu) \, (\rho \,T +  \langle \rho' \, \theta \rangle)$, whereas the equation for the pressure fluctuation is $p = (\kappa_b / m_\mu) \, (\rho \,\theta + \rho' \,T + \rho' \, \theta - \langle \rho' \, \theta \rangle)$. Here $\kappa_b$ is the Boltzmann constant and $m_\mu$ is the mass of molecules of surrounding fluid. Nonlinear terms $\rho' \, \theta$ and $\langle \rho' \, \theta \rangle$ in this equations are of the order of $O(M^4)$ [see, e.g., Chassaing et al., 2002], and for fluid flows with small Mach numbers, they can be neglected. This yields Eq.~(\ref{DS7}) for the ratio $p/P$.

In order to demonstrate that the directions of the mean flux of particles and the turbulent heat flux of temperature coincide, let us consider fluid with the mean temperature gradient (Fig.~1). Assume that the mean temperature $T_2$  at point 2 is larger than the mean  temperature $T_1$  at point 1. Let us consider two small volumes $a$ and $b$ located between these two points and let the direction of the local turbulent velocity of the volume $a$ at some instant be the same as the direction of the turbulent heat flux, $\langle {\bf u} \, \theta \rangle$, i.e., towards the point 1. Let the local turbulent velocity of the volume $b$ be directed at this instant opposite to the turbulent heat flux (i.e., to the point 2). In a fluid with an imposed mean temperature gradient, fluctuations of temperature $\theta$  and velocity ${\bf u}$ are correlated. Positive temperature fluctuations result in
positive pressure fluctuations. Consequently, the fluctuations of the temperature $\theta$ and pressure $p$  are positive inside the volume $a$ and negative inside the volume $b$. The fluctuations of the particle number density $n$  are positive in the control volume $a$ (because particles are locally accumulated in the vicinity of the maximum of pressure), and they are negative at the volume $b$ (because there is an outflow of particles from regions with a low pressure). Consequently, the mean flux of particles is positive in the volume $a$ (i.e., it is directed to the point 1), and it is also positive inside the volume $b$ (because fluctuations of velocity and number density of particles are negative in the volume $b$). Therefore, the mean flux of particles is non-zero and directed, as well as the turbulent heat flux $\langle {\bf u} \, \theta \rangle$ of temperature, towards the point 1.

\begin{figure}
\vspace*{10mm} \centering
\includegraphics[width=8cm]{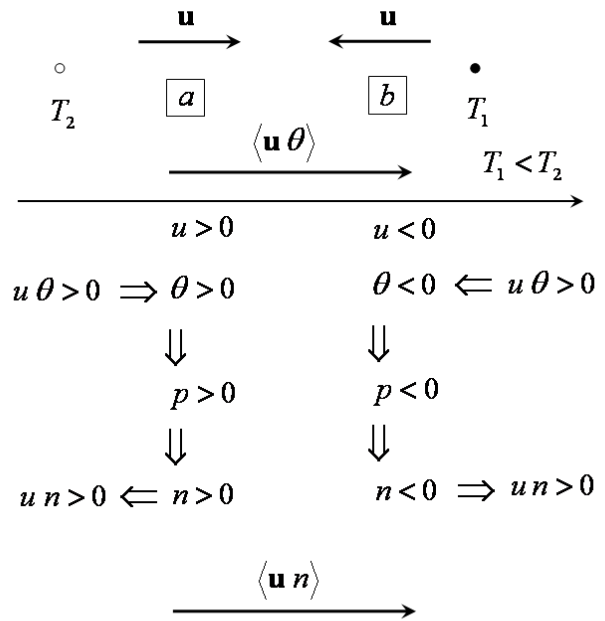}
\caption{\label{FIG-1} Mechanism of turbulent thermal diffusion.}
\end{figure}

\subsection{Turbulent flux of aerosols in stratified fluid}

The theory of turbulent thermal diffusion developed previously [see Elperin et al., 1996, 1997a, 1998, 2000c, 2001], does not take into account the effects of the Richardson number and anisotropy of turbulence. In this study, the parameters of turbulent thermal diffusion are derived as functions of the flux Richardson number and other turbulence characteristics for stably stratified turbulent flows. Equation for the evolution of the mean number density $N$ of particles reads:
\begin{eqnarray}
{\partial N \over \partial t} + \bec\nabla {\bf \cdot} \, \big[N \, ({\bf U} + {\bf W}_{g}) + {\bf F}^{(n)} \big] =0 \;,
\label{BB4}
\end{eqnarray}
where ${\bf U} =(U_1, U_2, U_3)$  is the mean fluid velocity (e.g., the wind velocity), ${\bf W}_{g} = \tau_s \, {\bf g}$  is the terminal fall velocity of particles, ${\bf g}$ is the acceleration of gravity. The equation for the turbulent flux of particles, ${\bf F}^{(n)} = \langle {\bf v} \, n \rangle$, is derived in Appendixes A, B, C using different approaches. The vertical component of the particle turbulent flux $F_z^{(n)}$ includes contributions of turbulent diffusion and turbulent thermal diffusion:
\begin{eqnarray}
F_z^{(n)} = V^{\rm eff}_z \, N - K_D \,\nabla_z N \;,
\label{BB5}
\end{eqnarray}
where $K_D$ is the coefficient of turbulent diffusion and $\nabla_z \equiv \partial /\partial z$. The effective velocity ${\bf V}^{\rm eff}$ caused by TTD is given by the following equation:
\begin{eqnarray}
{\bf V}^{\rm eff} = - t_T \, \langle {\bf v} \, \bec\nabla {\bf \cdot} \, {\bf v} \rangle \;,
\label{BB6}
\end{eqnarray}
where $t_T = \ell / E_K^{1/2}$  is the turbulent time,  $\ell$  is the turbulent length scale, $E_K$  is the turbulent kinetic energy. Equation~(\ref{BB6}) for the effective velocity has been derived using different rigorous methods by Elperin et al. [1996, 1997a, 1998, 2000c, 2001], Pandya and Mashayek [2002] and Reeks [2005]. Note that even a simple dimensional analysis yields the estimate for the effective velocity ${\bf V}^{\rm eff}$ that coincides with Eq.~(\ref{BB6}). Indeed, let us average Eq.~(\ref{BB1}) over the ensemble of the turbulent velocity field and subtract the obtained averaged equation from Eq.~(\ref{BB1}). This yields equation for the fluctuations $n$ of particle number density
\begin{eqnarray}
{\partial n \over \partial t} + \bec\nabla {\bf \cdot} \, (N \, {\bf
v} + {\bf Q}) = D \triangle n \;,
\label{BB7}
\end{eqnarray}
where ${\bf Q} = {\bf v} n - \langle {\bf v} n \rangle$  is the nonlinear term. The magnitude of $\partial n / \partial t + \bec\nabla {\bf \cdot} \, {\bf Q} - D \bec{\nabla}^2 n$ can be estimated as $n/t_T$. Therefore, the turbulent component $n$  of particle number density is of the order of  $n \approx - t_T \, \bec\nabla {\bf \cdot} \, (N \, {\bf v}) = - t_T \, [N \bec\nabla {\bf \cdot} \, {\bf v} +  ({\bf v} {\bf \cdot} \bec{\nabla}) N]$. Now let us calculate the turbulent flux of particles $F_i^{(n)} = \langle v_i \, n \rangle$:
\begin{eqnarray}
F_i^{(n)} = - N \, t_T \, \langle v_i \, \bec\nabla {\bf \cdot} \, {\bf v} \rangle  - t_T \, \langle v_i v_j \rangle \nabla_j N \;,
\label{BB8}
\end{eqnarray}
where the first term in the right hand side of Eq.~(\ref{BB8}) determines the turbulent flux of particles caused by turbulent thermal diffusion:  $- N \, t_T \, \langle v_i \, \bec\nabla {\bf \cdot} \, {\bf v} \rangle = V^{\rm eff}_i \, N$, while the second term in the right hand side of Eq.~(\ref{BB8}) determines the turbulent flux of particles caused by turbulent diffusion: $t_T \, \langle v_i v_j \rangle \nabla_j N = K_D \,\nabla_i N$. In the latter estimate we neglected the anisotropy of turbulence for simplicity.

For non-inertial particles advected by a turbulent fluid flow,
particle velocity ${\bf v}$ coincides with fluid velocity ${\bf u}$, and $\bec{\nabla} {\bf \cdot} {\bf u} \approx - ({\bf u}
{\bf \cdot} \bec{\nabla}) \rho / \rho \approx ({\bf u} \cdot
\bec{\nabla}) T / T$, where $ \rho $ and $T$ are the density and
temperature of the fluid. This formula takes into account the equation of state for an ideal gas but neglects small gradients of the mean fluid pressure, i.e., $(\nabla_z \rho) / \rho \approx - (\nabla_z T) / T$. Thus, the effective velocity~(\ref{BB6}) for non-inertial particles is determined by the following equation:
\begin{eqnarray}
V^{\rm eff}_z = - K_D \, {\nabla_z T \over T}
\; .
\label{BB6b}
\end{eqnarray}
Alternative derivations of Eqs.~(\ref{BB5}) and~(\ref{BB6b}) for non-inertial particles are also presented in Appendixes A and B.

\subsection{Effective velocity  of aerosols caused by turbulent thermal diffusion}

Let us now consider inertial particles and determine the dependence of the vertical component of the effective velocity $V^{\rm eff}_z$ on parameters characterizing  turbulence, the mean temperature profiles and particles. Note that the deviation of the particle velocity ${\bf v}$  from the fluid velocity ${\bf u}$  is small, but the deviation of $\bec\nabla {\bf \cdot} \, {\bf v}$ from   $\bec\nabla {\bf \cdot} \, {\bf u}$ is not small (see Eq.~(\ref{BB3})). Equations~(\ref{BB3}) and~(\ref{BB6}) yield
\begin{eqnarray}
{\bf V}^{\rm eff} = - K_D \, {\nabla_z T \over T} - {t_T \, \tau_s \over \rho} \langle {\bf v} \, \,\bec{\nabla}^2 p \rangle \;,
\label{BB6c}
\end{eqnarray}
where the second term in Eqs.~(\ref{BB6c}) describes the contribution of the particle inertia effect to the effective velocity. In order to determine this contribution we use the following formulae:
\begin{eqnarray}
&& {\tau_s \over \rho} \, \langle u_z \, \bec{\nabla}^2 p \rangle = {W_g L_P \over T} \, \langle u_z \, \bec{\nabla}^2 \theta \rangle \;,
\label{I1}\\
&& \langle u_z \, \bec{\nabla}^2 \theta \rangle = {2 \over 3  \, t_T} \,
\ln({\rm Re}) \, B({\rm Re}, a_\ast) \, \nabla_z T \;,
\label{I2}
\end{eqnarray}
[see  Elperin et al., 1996, 1998, 2000a], where $\theta$ are fluctuations of the temperature, $T$  is the mean temperature, $L_P^{-1} = |\nabla_z P / P|$ and $P$ is the mean fluid pressure, ${\rm Re} = \ell \, E_K^{1/2} / \nu$ is the Reynolds number, $\nu$  is the kinematic viscosity, and  $a_\ast$ is the particle size (e.g., the particle diameter). In Eqs.~(\ref{BB6c}) and~(\ref{I1}) we neglected small effects $\sim {\rm O}(\tau_s^2)$. For derivation of Eq.~(\ref{I1}) we took into account the equation of state, neglected the flux of fluid mass $\langle {\bf u} \, \rho' \rangle$ for the low-Mach-number flows, and used the identity $\tau_s = \rho \, W_g \, L_P / P$, where $|\nabla_z P| = \rho \, g$. Derivation of Eq.~(\ref{I2}) is given in Appendix C.

When the particle size  $a_{\ast} < a_{\rm cr}$, the function $B({\rm Re}, a_{\ast}) = 1$, and for $a_{\ast} \geq a_{\rm cr}$  the function $B({\rm Re}, a_{\ast})$  is given by $B({\rm Re}, a_{\ast}) = 1 - 3 \ln(a_{\ast} / a_{\rm cr}) / \ln({\rm Re})$  [see Elperin et al., 2000a], where the critical particle size is $a_{\rm cr} = \ell_{\nu} (\rho / \rho_{p})^{1/2}$, and $\ell_{\nu} = \ell \, {\rm Re}^{-3/4} $ is the Kolmogorov viscous scale of turbulence. The vertical component of the effective velocity  caused by turbulent thermal diffusion can be rewritten as follows:
\begin{eqnarray}
V^{\rm eff}_z &=& K_D \, {\nabla_z \rho \over \rho} - K_{TD} \, {\nabla_z T \over T} = - \alpha_{_{TD}} \, K_D \, {\nabla_z T \over T}\;,
\nonumber\\
\label{BB9}
\end{eqnarray}
where $\alpha_{_{TD}}=1 + K_{TD} / K_{D}$.
In order to determine the explicit form of the coefficient of turbulent diffusion $K_{D}$  and the coefficient $K_{TD}$  in Eq.~(\ref{BB9}), we have to use some turbulent closure model. In this study we use the turbulent closure model for stably stratified flows developed by Zilitinkevich et al. [2007]. This model is based on the budget equations for the key second moments: turbulent kinetic and potential energies, and vertical turbulent fluxes of momentum and buoyancy (proportional to potential temperature). This model takes into account the non-gradient correction to the traditional buoyancy flux formulation and implies the existence of turbulence at any gradient Richardson number. Predictions from this model are consistent with the available data from atmospheric and laboratory experiments, direct numerical simulation and large-eddy simulation.

The turbulent closure model by Zilitinkevich et al. [2007] is developed in the geophysical approximation for stably stratified flow, whereby the vertical mean fluid velocity, $U_3$, is negligibly small compared to the horizontal velocities, $U_1$  and $U_2$. A useful approximation for stably stratified flows is that the horizontal derivatives of the mean velocity components $U_{1,2}$ are negligibly small compared to their vertical derivatives. An effect of the horizontal derivatives of the mean velocity components $U_{1,2}$ on turbulent transport of particles in stably stratified flows is a subject of a separate study.

Using the turbulent closure model by Zilitinkevich et al. [2007] we obtain formulas for the coefficient of turbulent diffusion $K_{D}$:
\begin{eqnarray}
K_D &=& 2 C_n \, E_z^{1/2} \, \ell_z \,\Big[1 - C_D \, {{\rm Ri}_f \over 2 C_K \, A_z \, (1-{\rm Ri}_f)} \Big]
 \nonumber\\
&& \times \Big[1 + C_D \, C_n {{\rm Ri} \over \hat E_z} \Big]^{-1} \;,
\label{BB10}
\end{eqnarray}
and for the coefficient $K_{TD}$ in Eq.~(\ref{BB9}) that accounts for particle inertia:
\begin{eqnarray}
K_{TD} &=& {2 \over 3} C_n \, W_g \, L_P \,
\ln({\rm Re}) \, B({\rm Re}, a_\ast)
\nonumber\\
&& \times \Big[1 + C_D \, C_n {{\rm Ri} \over \hat E_z} \Big]^{-1} \; .
\label{BB11}
\end{eqnarray}
Derivation of Eqs.~(\ref{BB10}) and ~(\ref{BB11}) is given in Appendixes C and D. In the above equations, $A_z = E_z /E_K$  is the vertical anisotropy of turbulence and $E_z$  is the vertical turbulent kinetic energy. The gradient Richardson number is defined as ${\rm Ri} = {\cal N}^2 / S^2$,  where ${\cal N}^2 = \beta \,\partial \Theta / \partial z$ is the squared Brunt-V\"{a}is\"{a}l\"{a} frequency, $S = \big[(\partial U_1 / \partial z)^2 + (\partial U_2 / \partial z)^2 \big]^{1/2}$ is the mean velocity shear, $\Theta$  is the mean potential temperature (or the mean virtual potential temperature accounting for specific humidity), that is defined as $\Theta = T (P_\ast / P)^{1-\gamma^{-1}}$. Here $P_\ast$ is the reference value of the mean fluid pressure $P$, $\, \gamma = c_p/c_v=1.41$ is the specific heat ratio, $\beta=g/T_\ast$  is the buoyancy parameter and $T_\ast$ is a reference value of the mean temperature. The flux Richardson number is defined as ${\rm Ri}_f = - \beta F_z / K_M S^2$, where $F_i = \langle u_i \theta_p \rangle$ is the flux of the potential temperature, $\theta_p$ are the fluctuations of the potential temperature, $K_M$  is the eddy viscosity, $\hat E_z =E_z /(S \ell_z)^2$, $\,\ell_z$ is the vertical turbulent length scale, $C_n$, $\,C_K$ and  $C_D$ are empirical dimensionless coefficients.

The turbulent closure model suggested by Zilitinkevich et al. [2007] yields the following formulas for the parameters in Eqs.~(\ref{BB5}), (\ref{BB9})-(\ref{BB11}):  the vertical anisotropy of turbulence  $A_z$,
\begin{eqnarray}
A_z = {\hat E_z \over 2 C_K \, \Psi_\tau \, (1 - {\rm Ri}_f)} \;,
\label{BB12}
\end{eqnarray}
the vertical turbulent length scale,
\begin{eqnarray}
\ell_z = z \, \Big[1 - {{\rm Ri}_f \over {\rm Ri}_f^\infty} \Big]^{4/3} \;,
\label{BB13}
\end{eqnarray}
where $\Psi_\tau = 0.2 \, (1 - {\rm Ri}_f)$,
\begin{eqnarray}
\hat E_z = {2 C_K \, \Psi_\tau  \over 3 (1+C_r)} \, \big[C_r \, \Psi_3 - \big(C_r \, \Psi_3 +3 \big) {\rm Ri}_f \big] \;,
\label{BB14}
\end{eqnarray}
and $\Psi_3 = 1 - C_3 \, {\rm Ri}_f$. For very large gradient Richardson numbers, the flux Richardson number ${\rm Ri}_f \to {\rm Ri}_f^\infty = 0.2$, and the function $\hat E_z \to \hat E_z^\infty = 2 C_\theta \, C_K \, \Psi_\tau^\infty \, {\rm Ri}_f^\infty$, where $C_\theta$, $\,C_3$  and  $C_r$ are empirical dimensionless coefficients. Since the coefficient of turbulent diffusion $K_D$ should be positive, the empirical constant $C_D < 2/3$. The empirical constants have been determined by comparing results from the local closure model by Zilitinkevich et al. [2007] with data from laboratory and field experiments, large-eddy simulations (LES) and direct numerical simulations (DNS): $C_3 = 2.3$,  $C_r = 3$,  $C_K = 1.1$,  $C_\theta = 0.3$, $C_n = 0.3$  and  $C_D = 0.3$.

The horizontal components of the particle flux are given by the following formulas:
\begin{eqnarray}
F_i^{(n)} = - C_n \,  {\ell_z \, (\nabla_z U_i) \over E_z^{1/2}}  \, F_z^{(n)} = - C_n \,  {F_z^{(n)} \over \hat E_z^{1/2}} \;,
\label{A9}
\end{eqnarray}
where $i =1,2$. Equation~(\ref{A9}) implies that in neutral stratification the horizontal turbulent flux of particles is of the same order as  $F_z^{(n)}$, whereas in strongly stable stratification it is approximately by a factor of  $z/L$ larger then  $F_z^{(n)}$. This flux generally deviates from the mean wind direction, thus contributing to the horizontal cross-wind dispersion.

The steady-state solution of Eq.~(\ref{BB4}) reads:
\begin{eqnarray}
{\nabla_z N \over N} + \alpha_{_{TD}} \, {\nabla_z T \over T} + {W_g \over K_D} = 0 \;,
\label{BB15}
\end{eqnarray}
where the turbulent thermal diffusion ratio $\alpha_{_{TD}}$ is given by the following equation:
\begin{eqnarray}
\alpha_{_{TD}}  &\equiv& 1 + {K_{TD} \over K_D} =
1 + {W_g \, L_P \, \ln({\rm Re}) \, B({\rm Re}, a_\ast) \over 3 \, E_z^{1/2} \, \ell_z}
\nonumber\\
&& \times \Big[1 - C_D \, {{\rm Ri}_f \over 2 C_K \, A_z \, (1-{\rm Ri}_f)} \Big]^{-1} \; .
\label{BB16}
\end{eqnarray}
For gases and very small particles with negligible sedimentation velocity $\alpha_{_{TD}} =1$.

In the present study we have assumed that the coefficients $K_{D}$ and $K_{TD}$ are independent of the aerosol concentration because we consider the case $m_p N \ll \rho$, where $m_p$ is the particle mass. When this condition is not valid, nonlinear effects of aerosols on atmospheric turbulence should be taken into account. This is a subject of a separate study.

Note that there is another effect that is called ''turbophoresis'' and related to the particle inertia. The turbophoresis results in an additional mean particle velocity due to inhomogeneity of turbulence [see Caporaloni et al., 1975; Reeks, 1983].
The turbophoresis and turbulent thermal diffusion are totally different phenomena. Indeed, averaging Eq.~(\ref{BB2}) over fluctuations we obtain the mean particle  velocity:
\begin{eqnarray}
({\bf V}_{p})_{i} = {\bf U}_{i} - \tau_s {\partial {\bf U}_{i}
\over \partial t} - \tau_s \nabla_{j} \langle u_{i} u_{j} \rangle + \tau_s \langle u_{j} (\bec\nabla {\bf \cdot} {\bf u}) \rangle \; .
\label{DS2}
\end{eqnarray}
For example, in isotropic turbulence $\langle u_{i} u_{j} \rangle = (1/3) \langle {\bf u}^2 \rangle \delta_{ij} $, and the mean particle  velocity reads:
\begin{eqnarray}
{\bf V}_{p} = {\bf U} - \tau_s {\partial {\bf U}
\over \partial t} - {\tau_s \over 3} \bec{\nabla}
\langle {\bf u}^{2} \rangle + \tau_s \langle {\bf u} (\bec\nabla {\bf \cdot} {\bf u}) \rangle \; .
\label{DS3}
\end{eqnarray}
The third term in Eq.~(\ref{DS3}) describes the effect of
turbophoresis due to inhomogeneity of turbulence. The ratio of the mean particle velocity due to turbophoresis to the effective particle velocity caused by the phenomenon of turbulent thermal diffusion is
\begin{eqnarray}
{|V^{\rm turbo}| \over |V^{\rm eff}_z|}  = {\tau_s \over t_{_{T}}} {|\bec{\bf \nabla} \langle {\bf u}^{2} \rangle| / \langle {\bf u}^{2} \rangle \over \alpha_{_{TD}} |\bec{\bf \nabla} T| / T} \; .
\label{DS4}
\end{eqnarray}
Since $\tau_s \ll t_{_{T}}$, the mean particle velocity due to turbophoresis is much smaller than the effective particle velocity caused by the phenomenon of turbulent thermal diffusion. Although both effects are related to the particle inertia, the effective particle velocity due to the phenomenon of turbulent thermal diffusion, originates from the turbulent particle flux $\langle {\bf u} \, n \rangle$, i.e., describes the collective statistical phenomenon, while the mean particle velocity due to turbophoresis originates directly from the expression for mean particle velocity.

The mechanism of phenomenon of turbulent thermal diffusion is also principally different from molecular thermophoresis (and molecular thermal diffusion). The basic difference between these phenomena is explained in the following. The phenomenon of turbulent thermal diffusion occurs due to a combined action of turbulence effects and particle inertia effect, while molecular thermophoresis is caused by purely kinetic effects related to thermal motion of molecules.
In stratified turbulent flow turbulent thermal diffusion and molecular thermophoresis occurs simultaneously, although the effect of turbulent thermal diffusion for large Reynolds numbers is essentially stronger than the effect of molecular thermophoresis. In particular, the ratio of the effective velocity due to turbulent thermal diffusion to the velocity caused by molecular thermophoresis is of the order of Reynolds number. In this estimate we use formula for the molecular thermophoretic velocity $V_{\rm th} \sim \nu \, |\bec{\nabla} T| / T$ [see Friedlander, 2000] and Eq.~(\ref{BB9}) for the effective velocity caused by turbulent thermal diffusion. For example, in the atmospheric turbulent boundary layer Reynolds numbers are of the order of $10^7$, that implies that molecular thermophoresis is negligibly small.

Note that in the context of thermal convection or stably stratified turbulence, an anelastic approximation $[\bec\nabla {\bf \cdot} (\rho \, {\bf u}) = 0]$ is used for low Mach numbers. This is standard for describing deep convection but introduces the possibility of a compressible fluid velocity tied to variations in the fluid density. The fluid mass flux is divergence-free but a tracer particle moves with the fluid velocity. The flow compressibility is linked to the temperature fluctuations and there are dynamic correlations between fluid density, temperature and pressure. Irrespective of particle inertia, this introduces a possibility of a mean drift even in homogeneous turbulence, i.e., causes turbulent thermal diffusion.

In the next sections we discuss the potential impact of the phenomenon of turbulent thermal diffusion on distribution of the atmospheric constituents. In particular, we analyze the observational information in the vicinity of strong temperature gradients (near the tropopause) in order to find out whether the effect of turbulent thermal diffusion is: (i) observable, (ii) significant, (iii) provides well-grounded explanations of the observed aerosol layers and their positions. For the lower troposphere, we use the numerical simulations as a tool to study the contribution of the effect of turbulent thermal diffusion to the lower-troposphere vertical profiles of aerosol concentration.
Note also that the temperature minimum in the atmosphere exists only for the absolute temperature while the potential temperature increases with altitude in the upper troposphere and the lower stratosphere.

\section{Aerosol measurements by GOMOS near the tropopause}

\subsection{Outline of the observational technique}

Tropopause is a well-known region with strong gradients of temperature and also with substantial amount of particles, which remain there over long periods. Simultaneous observations of the vertically-resolved aerosol concentrations and temperature are scarce but the presence of the aerosol layers in the vicinity of the tropopause is well established [Brasseur and Solomon, 2005].

For joint analysis of temperature profiles and aerosol concentrations, we use a unique dataset obtained from GOMOS instrument (Global Ozone Monitoring by Occultation of Stars) onboard the Envisat satellite [Kyr\"{o}l\"{a} et al., 2004; Bertaux et al., 2004, http://envisat.esa.int/instruments/gomos]. GOMOS is equipped with the UV/Visible/NIR spectrometers, which record stellar spectra transmitted through the atmosphere continuously as the star sets behind the Earth limb. The measurements are performed in the limb-viewing geometry with the sampling frequency of 2 Hz. The atmospheric transmission spectra obtained after dividing the stellar spectra observed through the Earth atmosphere by the reference spectrum, recorded above the atmosphere, contain spectral features of absorption and scattering by gases and particles. This allows reconstructing the vertical profiles of O$_3$, NO$_2$, NO$_3$, O$_2$, H$_2$O and aerosol extinction in the atmosphere. The vertical sampling resolution of GOMOS data is 0.5-1.7 km.

While ozone can be retrieved up to 100 km altitude, other species are usually detectable in the upper troposphere and in the stratosphere. The lowest altitude of GOMOS measurements is from ~5 km to ~20 km; it depends mainly on stellar brightness and clouds top height. The stellar light can be transmitted only through thin clouds (like subvisual cirrus clouds), and they appear as increased aerosol extinction in GOMOS data. Usually, aerosols are retrieved with a good accuracy down to tropopause and slightly below from occultations of bright stars.

The GOMOS inversion of the chemical composition is performed in two steps [Kyr\"{o}l\"{a} et al., 1993]. First, atmospheric transmission data from every tangent height are inverted to horizontal column densities (along the path of the light beam from the star) for gases and optical thickness for aerosols (spectral inversion). Then, for every constituent, the collection of the horizontal column densities at successive tangent heights is inverted to vertical density profiles (vertical inversion).

Since the aerosol extinction spectrum is not known a priori, a second-degree polynomial model is used for the description of the aerosol extinction $\beta_{\rm aero}$  in the GOMOS retrievals:
\begin{eqnarray}
\beta_{\rm aero} &=& {\sigma_0 \over \lambda} \, N(z) \big[1 + c_1 (\lambda - \lambda_{\rm ref}) + c_2 (\lambda - \lambda_{\rm ref})^2\big] \;,
\nonumber\\
\label{BB17}
\end{eqnarray}
where $\lambda$ is wavelength in nm,  $\lambda_{\rm ref} =500$ nm, $\sigma_0 = 3 \times 10^{-7}$ cm$^2$ nm is the scaling factor,  $N(z)$  is the aerosol mean number density at altitude $z$ (in the units of  cm$^{-3}$), and parameters $c_1$ and $c_2$ determine the wavelength dependence of the aerosols extinction spectra. The aerosol number density $N$ and the parameters $c_1$ and $c_2$  are retrieved from GOMOS data.

High-resolution temperature profiles (HRTP) in the stratosphere and the upper troposphere are retrieved from the synchronous scintillation measurements by the GOMOS fast photometers operating at 1 kHz sampling frequency at red (650-700 nm) and blue (470-520 nm) wavelengths [Dalaudier et al., 2006]. The measurement principle exploits the chromatic refraction in the atmosphere. The bi-chromatic scintillations recorded by the photometers allow accurate determination of a refractive angle, which is proportional to the time delay between the photometer signals. The procedure of conversion of refractive angle profiles to temperature profiles is similar to that used in radio occultation. At altitudes ~18-35 km, HRTP is retrieved with the vertical resolution 200-250 m and the accuracy of 1-2 K. The best accuracy is achieved in vertical (in orbital plane) occultations of bright stars [Sofieva et al., 2007]. Below ~15 km, the quality of HRTP decreases due to low signal-to-noise ratio, broadening of scintillation peaks as a result of chromatic smoothing and the violation of the assumptions used in retrievals (in particular, the weak scintillation assumption).

\subsection{Observed aerosol and temperature profiles}

For joint analysis of HRTP and aerosol profiles, two data-sets were selected:  217 successive occultations of one of the brightest stars, Canopus (referred hereafter as to S002), with visual magnitude -0.7 and the effective temperature 7,000 K corresponding to location of the ray perigee point over 36 N-37 N (mid-latitudes) in Jan-Feb 2003, and 560 successive occultations of S029 (visual magnitude 1.6, the effective temperature 10,200 K) located at 10-20 N, for the same period. The observation period was selected arbitrarily. Random inspections of aerosol and temperature profiles at different locations and seasons at low and mid latitudes have indicated that the features described below are common.

Since the vertical resolution of HRTP (250 m) is much finer than that of aerosol profiles (2 km), HRTP were smoothed down to resolution of aerosol profiles. For the sake of reliability, we also included the temperature profiles extracted from archive of operational analysis of European Centre of Medium-Range Weather Forecast (ECMWF). Usually, smoothed HRTP and ECMWF profiles are in a very good agreement in the considered altitude range.

It has been noticed that the aerosol concentration and temperature profiles are often anti-correlated (see a few examples in Figures~2 and~3). At a qualitative level, this is in agreement with the theoretical predictions related to the impact of the phenomenon of turbulent thermal diffusion on the aerosol profiles (the impact of other effects is considered in the discussion section).

\begin{figure}
\vspace*{8mm} \centering
\includegraphics[width=8cm]{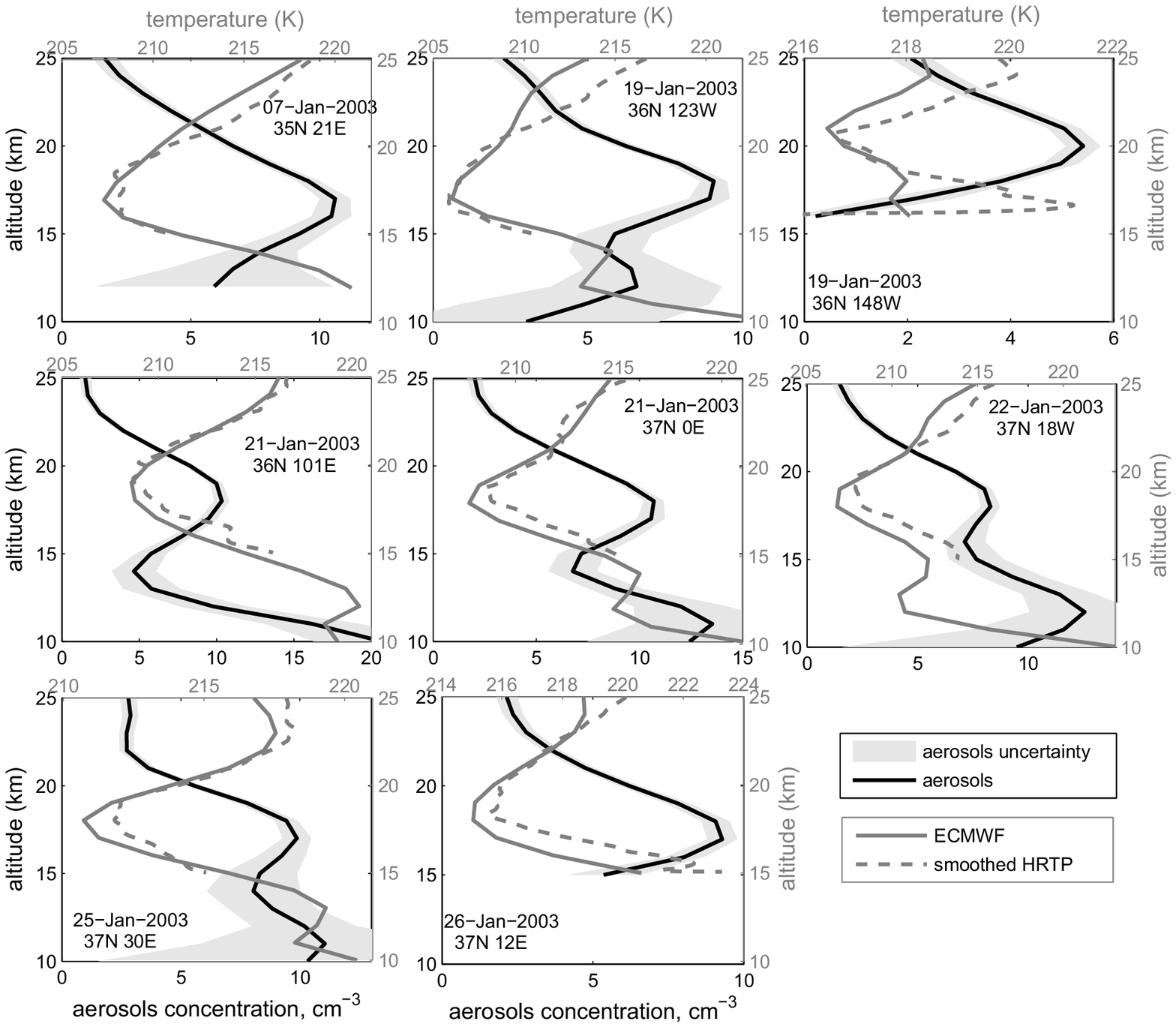}
\caption{\label{FIG-2} Examples of aerosols (black lines) and temperature profiles (dashed grey lines) observed by GOMOS for middle latitudes (36-37N) in January 2003, and the independent ECMWF temperature analysis data at measurement locations (grey solid lines).}
\end{figure}

\begin{figure}
\vspace*{8mm} \centering
\includegraphics[width=8cm]{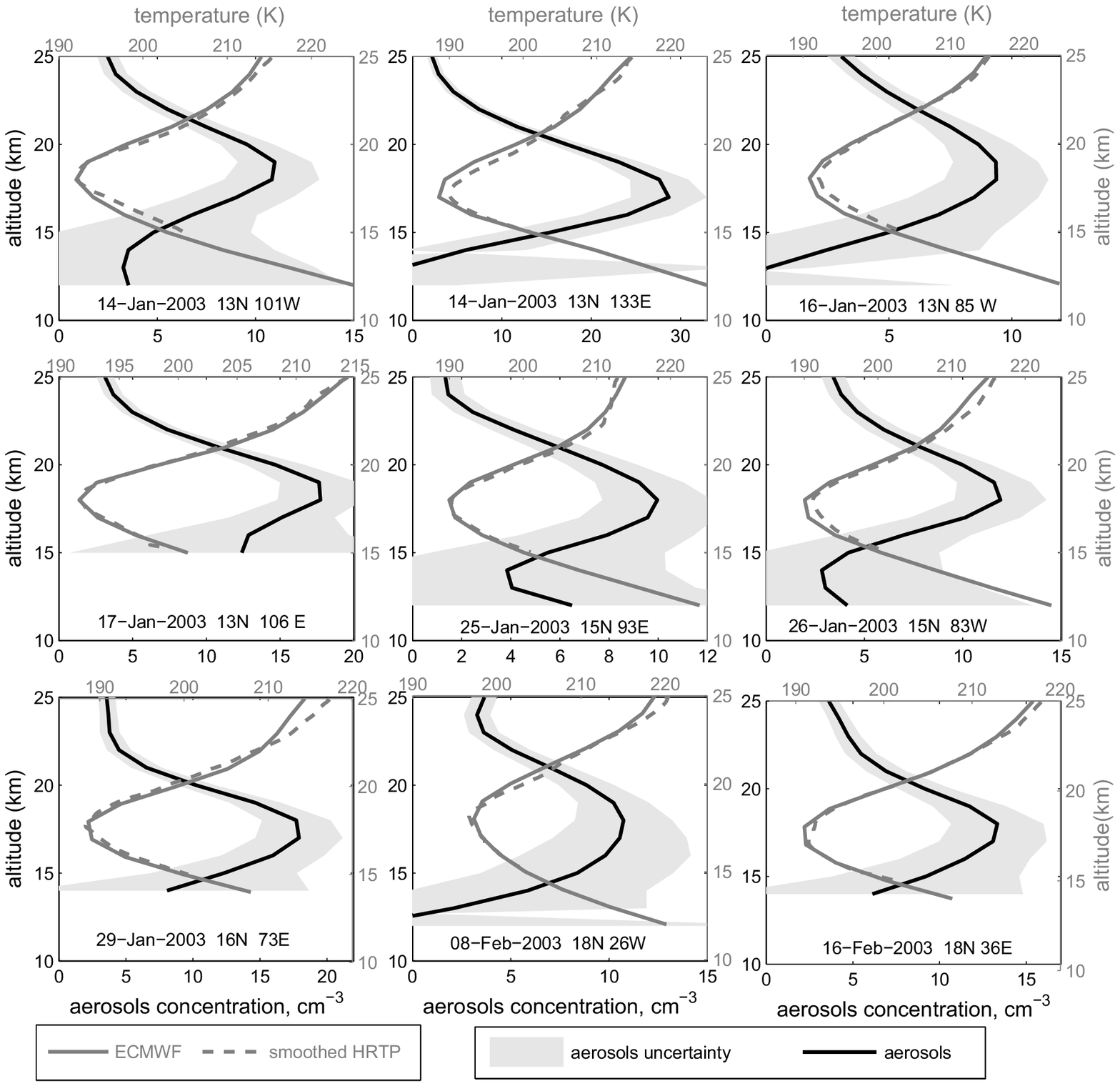}
\caption{\label{FIG-3} As Figure 2, but for locations over tropics.  All notations are as in Figure 2.}
\end{figure}

Similarities between the results in the low- and mid-latitude regions suggest a common mechanism behind the observed anti-correlation. The data for the equatorial tropopause, however, have to be treated with care: very low temperatures ($T < 198$ K) at the tropical tropopause are favorable for formation of cirrus clouds [Brasseur and Solomon, 2005], which are seen as increased aerosols extinction in the GOMOS data.
Nevertheless, current understanding and observations of the cirrus clouds position them slightly below the tropopause, while in our cases the aerosol layer is practically symmetrical with regard to the temperature profile (unfortunately, the vertical resolution was insufficient for unambiguous conclusions). It must be noted that the cirrus cloud formation is not important for mid-latitude profiles, because the observed temperature over 206 K is too high for the formation of the cirrus clouds. The latter usually occurs for the temperature smaller than 198 K (Jensen et al., 1996).

The slope of aerosol extinction spectra can serve as an indicator of the aerosol type: large particle have flat spectra in UV/Visible range, while small particles have larger extinction in UV than in Visible range, which approaches the $\lambda^{-4}$ scattering law in case of very small particles (see, e.g., discussion by Vanhellemont et al. [2005]).  The analysis of the slope of the GOMOS aerosol extinction spectra also indicates that the observed mid-latitude aerosols are probably background sulfate aerosols (small particles) rather than ice crystals (large particles).

The statistically significant anti-correlation between the temperature and the aerosol concentration has been observed for 30 \% profiles at mid-latitudes and for 50 \% of profiles in tropics (see Figure~4). The distributions of the correlation coefficient between the relative gradients of temperature and the aerosol concentration are strongly skewed with modes close to $-1$.

\begin{figure}
\vspace*{8mm} \centering
\includegraphics[width=8cm]{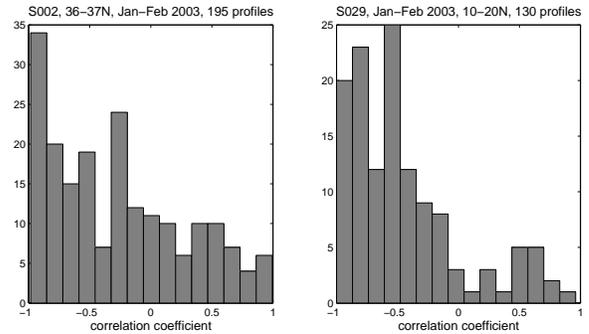}
\caption{\label{FIG-4} Histogram of correlation coefficients between the ECMWF temperature profiles and GOMOS aerosol retrievals. Left: occultations of Canopus, mid latitudes (195 profiles out of 217 total covering the tropopause region), right: occultations of S029, tropics (130 profiles out of 560 covering the tropopause region). For rejected profiles the retrievals have been terminated above the tropopause due to dense clouds.}
\end{figure}

A further insight can be obtained from quantitative analysis of the profiles and, first of all, the functional relation between temperature and aerosol concentrations. The theory of turbulent thermal diffusion predicts the following relation between the steady-state profiles of the mean temperature and mean particle number density:
\begin{eqnarray}
N(z) \, [T(Z)]^{\alpha_{_{TD}}} = \exp \Big[-\int_0^z \, {W_g \over K_D} \, dz' \Big] \;,
\label{BB18}
\end{eqnarray}
which follows from Eq.~(\ref{BB15}). Here the turbulent thermal diffusion ratio $\alpha_{_{TD}}$ is determined by Eq.~(\ref{BB16}). Note that Eq.~(\ref{BB18}) has been previously used by Eidelman et al. (2004, 2006a,b) for analysis of laboratory measurements of TTD.

Equation~(\ref{BB18}) allows experimental determination of $\alpha_{_{TD}}$ from the measured profiles of temperature and aerosol concentration without using Eq.~(\ref{BB16}). Indeed, assuming that the right-hand-side of Eq.~(\ref{BB18}) is constant above and below the temperature minimum (not necessarily the same constants), yields the power-law type regression with unknown constant parameter. Taking logarithm and differentiating Eq.~(\ref{BB18}) with respect to height, we arrive at a linear regression equation for determining  $\alpha_{_{TD}}$. Application of this procedure to GOMOS profiles with statistically significant negative correlation (at the significance level of 99 \%) yields the histograms of $\alpha_{_{TD}}$ shown in Figure 5, which have a pronounced peak and a narrow width. As one can see in Figure~5, the values of $\alpha_{_{TD}}$  for 60-70 \% of the anti-correlated profiles are very similar. This modal value, in turn, varies with latitudes being smaller for equatorial regions.

\begin{figure}
\vspace*{8mm} \centering
\includegraphics[width=8cm]{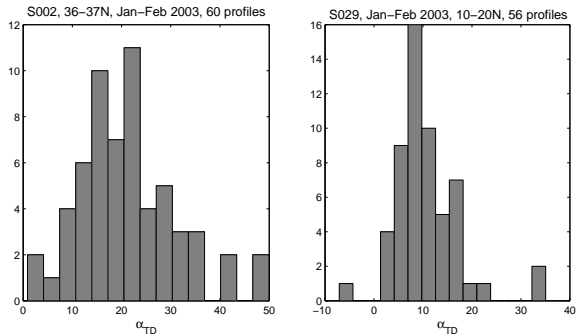}
\caption{\label{FIG-5} Histogram of the TTD ratio $\alpha_{_{TD}}$: a regression between the relative gradients of temperature and number concentrations of particles for profiles having statistically significant negative correlation. Left: mid latitudes (S002 measurements), right: tropics (S029 measurements).}
\end{figure}

In the comparison of the observed values of $\alpha_{_{TD}}$  with the theoretical predictions we have to take into account several parameters affecting it. The theoretical predictions for $\alpha_{_{TD}}$  [see Eq.~(\ref{BB16})] relate it to:

(i) atmospheric pressure that affects kinematic viscosity and the mean free path of molecules;

(ii) the coefficient of turbulent diffusion $K_D$;

(iii) aerosol size and material density.

Theoretical dependencies of the turbulent thermal diffusion ratio $\alpha_{_{TD}}$ versus the coefficient of turbulent diffusion $K_D$ are shown in Figs.~6-7, and the function $\alpha_{_{TD}}$ versus the size of aerosols is shown in Fig.~8. Theoretical values of $\alpha_{_{TD}}$ have been determined using Eq.~(\ref{BB16}) for ${\rm Ri}_f = 0$ and $L_P = 8$ km. Increasing ${\rm Ri}_f$ results in the increase of the turbulent thermal diffusion ratio $\alpha_{_{TD}}$. In calculations we used the following formulas for the Reynolds number, ${\rm Re} = 3 K_D / 2C_n \nu$, and for the Stokes time $\tau_s = C_c \, \rho_a \, a^2_\ast \, / (18 \, \rho \, \nu)$, where $C_c$ is the Cunningham correction factor.

\begin{figure}
\vspace*{8mm} \centering
\includegraphics[width=8cm]{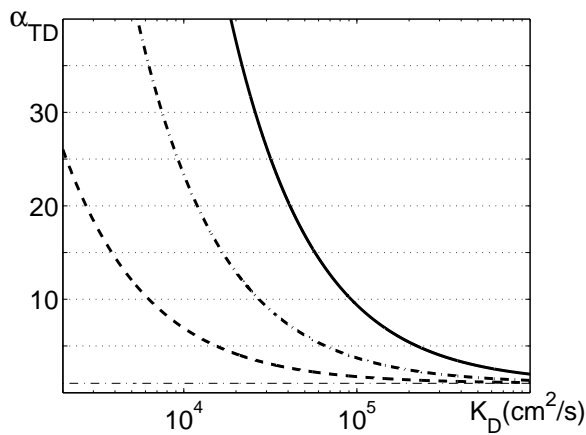}
\caption{\label{FIG-6} Theoretical values of the TTD ratio $\alpha_{_{TD}}$  versus the coefficient of turbulent diffusion $K_D$  for the altitude 17 km and different sizes of aerosols: 5  $\mu$m (solid line), 2.5  $\mu$m (dashed-dotted line) and 1  $\mu$m (dashed line). Thin dashed-dotted line $\alpha_{_{TD}}=1$ corresponds to non-inertial particles.}
\end{figure}

\begin{figure}
\vspace*{8mm} \centering
\includegraphics[width=8cm]{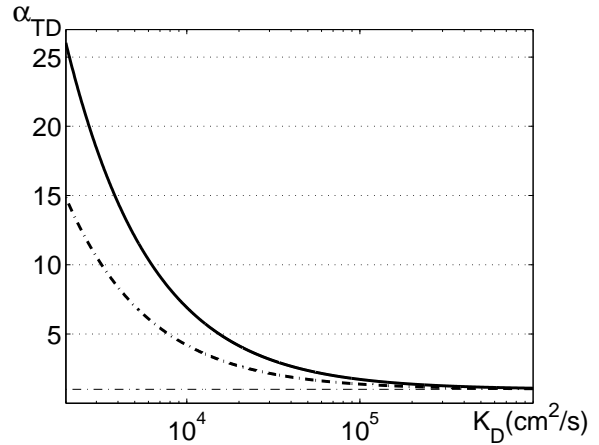}
\caption{\label{FIG-7} Theoretical values of the TTD ratio $\alpha_{_{TD}}$  versus the coefficient of turbulent diffusion $K_D$  for aerosols of 1 $\mu$m size and different altitudes: 17 km (solid line)  and 1 km (dashed-dotted line). Thin dashed-dotted line $\alpha_{_{TD}}=1$ corresponds to non-inertial particles.}
\end{figure}

\begin{figure}
\vspace*{8mm} \centering
\includegraphics[width=8cm]{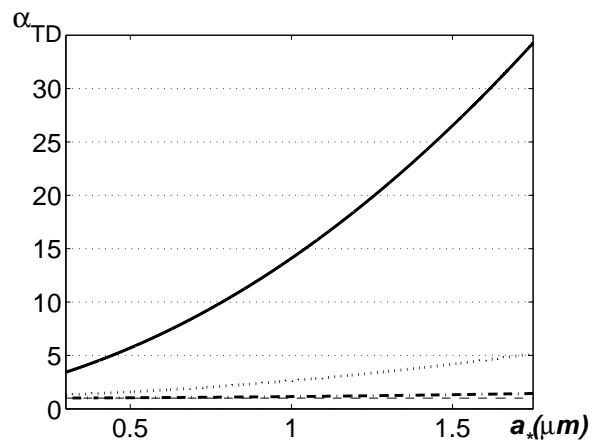}
\caption{\label{FIG-8} Theoretical values of the TTD ratio $\alpha_{_{TD}}$  versus the size of aerosols for $K_D = 3 \times 10^3$  cm$^2$  s$^{-1}$  and the altitude 18.1 km (solid line); $K_D = 3 \times 10^4$  cm$^2$  s$^{-1}$  and the altitude 1 km (dotted line); $K_D = 3 \times 10^5$  cm$^2$  s$^{-1}$  and the altitude 1 km (dashed-dotted line). Thin dashed line $\alpha_{_{TD}}=1$ corresponds to non-inertial particles.}
\end{figure}

Inspection of Figure~6 (the dependence of $\alpha_{_{TD}}$ at 17 km altitude versus $K_D$ and aerosol size) reveals that the variations of the turbulent thermal diffusion ratio $\alpha_{_{TD}}$ can be very large. The dashed line (aerosol of 1 $\mu$m in diameter) shows that $\alpha_{_{TD}}$  exceeds 7 when $K_D \sim 1$ m$^2$ s$^{-1}$ or less. The parameter $\alpha_{_{TD}}$ is much larger than those obtained in the laboratory studies [see Elperin et al., 2000b; Eidelman et al., 2004, 2006a, 2006b], which have been conducted for much higher values of $K_D$ and at surface pressure. The comparison with surface-pressure conditions for $1  \mu$m aerosols (Figure~7) shows that lower air density at the tropopause almost doubles the parameter $\alpha_{_{TD}}$, with another factor of 5-10 due to the lower eddy diffusivity. The combined impact of a lower fluid density and a smaller eddy diffusivity is illustrated in Figure 8, where the solid line corresponds to the tropopause conditions while other curves correspond to the case of the lower troposphere.

Comparison of the histograms in Figure~4 and Figures~6-8 demonstrates that the observed values of $\alpha_{_{TD}}$  are fairly close to the predicted ones. The theory of turbulent thermal diffusion also provides a natural explanation for the lower values of $\alpha_{_{TD}}$  observed in tropics. Despite the higher altitude and the lower pressure, the dominating parameter is the turbulent diffusion coefficient $K_D$,
which is often high in the equatorial region [Paramesvaran et al., 2003; Fujiwara et al., 2003; Yamamoto et al., 2003]. In fact, one can consider an inverse problem: the observed aerosol concentration and temperature profiles can be used for determining the turbulent diffusion coefficient $K_D$ in each particular case using $\alpha_{_{TD}}$  as an input parameter. This would yield the information about the level of turbulence near the tropopause obtained indirectly from the aerosol concentration and temperature profiles.

In this section we have shown that the effect of turbulent thermal diffusion can be significant even for small particles. Indeed, for 1-2 $\mu$m particles, the fluid velocity ${\bf u}$ is much larger than the inertial term $|\tau_s d {\bf u} / dt|$, and usually particles of this size are used in laboratory experiments as flow tracers. However, the phenomenon of turbulent thermal diffusion is determined by two contributions: (i) the correlation of the divergence of the inertial term, $\bec\nabla {\bf \cdot} \, (\tau_s d {\bf u} / dt)$, with turbulent velocity and (ii) the correlation of the divergence of the non-inertial term, $\bec\nabla {\bf \cdot} \, {\bf u}$, with turbulent velocity. This results in effective particle velocity that can be larger than the terminal fall velocity $ W_g$. Inspection of Figs. 6-7 shows that $\alpha_{_{TD}}$ can be much larger than 1 for $K_D \ll 10^6$ cm$^2 /$ s. Consequently, the contribution to the effective velocity of aerosols [see Eqs.~(\ref{BB9}) and (\ref{BB16})] caused by the inertia term is larger than the contribution of the non-inertial term. On the other hand, for the lower troposphere (see the next Section), whereby the turbulent diffusion coefficient $K_D \sim (1 -5) \times 10^5$ cm$^2 /$ s, the parameter $\alpha_{_{TD}} \approx 1$, and the effective velocity of aerosols is due to the non-inertial term.

\section{Assessment of the contribution of the effect of turbulent thermal diffusion to the lower troposphere composition}

Analysis of the role of turbulent thermal diffusion in the lower troposphere is quite involved due to highly dynamic character of the atmospheric boundary layer, in particular, its strong diurnal cycle. Varying wind and numerous sources injecting aerosols at various heights pose additional difficulties. With these conditions, observational evidence of the turbulent thermal diffusion in the lower troposphere is very difficult to detect, and a preliminary analysis of available radar data has not produced unequivocal results.

The influence of turbulent thermal diffusion in the troposphere was evaluated via numerical modelling. This section discusses the means of including the phenomenon of turbulent thermal diffusion into the chemical transport models and presents results of several numerical experiments with the modelling system SILAM in order to reveal the impact of formulations of turbulent thermal diffusion on the predicted aerosol distribution.

\subsection{Setup of the modelling experiment}

The SILAM system [Sofiev et al., 2006] is a dual-core Lagrangian-Eulerian dispersion model. Its meteorological pre-processor provides the system with all parameters required for the simulations, including the input for TTD assessment. For the experiments, we used the Eulerian dynamic core based on original advection scheme by Galperin [2000] with the vertical turbulent diffusion parameterized via K-theory and implemented in the model via extended resistive analogy by Sofiev [2002]. This vertical diffusion scheme meets the key requirement of the current experiment because it explicitly treats the particle sedimentation and allows virtually any profile of vertical turbulent mixing coefficient $K_D$.

The goal of the numerical experiment was to evaluate the impact of TTD on the aerosol concentrations in the lower troposphere. For that purpose, the model was run through a reference year 2000 over the European continent using the archive of the limited-area meteorological model HIRLAM [Unden et al., 2002]. As an aerosol tracer, we used primary anthropogenic particles (i.e. aerosols directly emitted from anthropogenic sources) in two size classes: fine particles with less than 2.5 $\mu$m in diameter - PM 2.5, and a coarse-mode fraction from 2.5  $\mu$m to 10  $\mu$m size - PM 2.5-10. To simplify the simulations, both particle size ranges were considered as single bins with characteristic diameters of 1.4  $\mu$m and 6  $\mu$m and densities of 1230 kg m$^{-3}$ and 1500 kg m$^{-3}$, respectively. Exchange of particles between these bins due to condensation and coagulation was neglected. Anthropogenic emission for both classes was taken from the EMEP database (European Monitoring and Evaluation Programme, http://www.emep.int).

The turbulent thermal diffusion velocity $V^{\rm eff}_z$ was introduced as an additive term to particle sedimentation velocity $W_g$ so that the total mean vertical particle velocity reads:
\begin{eqnarray}
V^{\rm tot} &=& W_g + V^{\rm eff}_z = {C_c \, \rho_a \, a^2_\ast \, g \over 18 \mu}  - K_D \, \alpha_{_{TD}} \, {\nabla_z T \over T} \;,
\nonumber\\
\label{BB19}
\end{eqnarray}
where $z$-axis is directed along the acceleration due to gravity ${\bf g}$,  $\, a_\ast$  is particle diameter, $C_c$ is the Cunningham correction factor and $\mu = \rho \nu$ is a dynamic viscosity of air.

Vertical turbulent diffusion coefficient $K_D$  is determined using SILAM standard meteorological processing routines, which evaluates $K_D$   in a simplified form avoiding the uncertain parameters, such as turbulent kinetic energy or turbulent length scale. The computations of $K_D$ start at the surface with evaluation of $K_D(z_1 = 1$ m) after Genikhovich et al. [2004] and the height of the boundary layer $H_{\rm ABL}$ as described by Sofiev et al. [2006]. Following the surface-layer assumptions, it is assumed that $K_D \propto z$  in the range from $z_1$  up to  $z_1 = 0.1 H_{\rm ABL}$, then $K_D$  remains constant up to the height $H_{\rm ABL}$. At the top of the boundary layer the coefficient $K_D$  sharply decreases by an order of magnitude and then remains constant up to the top of the modelled domain. These formulations are evidently too crude for detailed evaluation of the near-tropopause phenomena discussed in the previous section but are sufficient for studying the lower troposphere processes, which were in the focus of the modelling experiment.

The standard resistance-based diffusion term with turbulent and laminar-layer resistances represents the diffusion pathway of the aerosol dry deposition. For the studied particle size range, in typical conditions and without turbulent thermal diffusion, the standard diffusion term is much smaller than the gravitational sedimentation. However, in our simulations, this term has been included because the TTD velocity $V^{\rm eff}_z$ can partly or entirely outweigh the sedimentation velocity, so that the diffusion component may become important.

\subsection{Results of the simulations}

The long-term averaging of the obtained results highlighted the mean influence of the turbulent thermal diffusion: moderate in absolute values but systematic uplift of the aerosols. The driving force of this uplift is the decrease of temperature with increase of the altitude. Examples in Figure~9 for Norwegian mountains (panel a) and Arctic Ocean (panel b) show that, depending on the region elevation, the TTD impact can result in both increase and decrease of the near-surface concentrations. In the low-altitude regions and close to the sources, the near-surface concentrations decrease while over the elevated and mountain areas they grow (Figure~10). The pollution masses (aerosols), which are raised to higher altitudes by the TTD velocity  $V^{\rm eff}_z$, appeared to be transported over larger distances and at higher speed since the mean wind at the plume height is stronger and the dry deposition is weaker. Both effects increase the overall transport distance of aerosols. In turn, this resulted in wider distribution of aerosols, higher concentrations in remote areas and a larger fraction of mass leaving the source region, such as Europe.

\begin{figure}
\vspace*{8mm} \centering
\includegraphics[width=8cm]{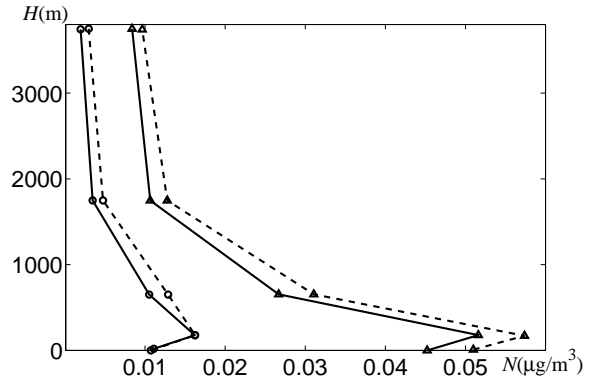}
\caption{\label{FIG-9} Mean annual vertical profiles with (dashed lines) and without (black lines) TTD for PM 2.5-10 over two characteristic regions: Norwegian mountains (triangles) and Arctic Ocean (circles). Unit is  $\mu$g m$^{-3}$.}
\end{figure}

\begin{figure}
\vspace*{8mm} \centering
\caption{\label{FIG-10} Mean annual near-surface concentrations of PM 2.5 and PM 2.5-10 without the TTD term (panels a and b, respectively), and a ratio between the PM 2.5 and PM 2.5-10 concentrations obtained with and without the TTD term in the SILAM model formulations (panels c and d, respectively). Unit of concentrations is $\mu$g m$^{-3}$.}
\end{figure}

The above changes are moderate in absolute values. In our numerical experiment, the transport of coarse particles PM 2.5-10 outside the modelled domain has increased by 5-15 \% depending on season. For PM 2.5 the difference was small. Episodically, the impact of TTD was more significant but the year-long simulations did not reveal any case when the daily concentrations changed by more than by the factor of 1.5 due to the TTD term.

The difference between the particle size sections can be explained as follows (see Figures~6 and~7). In the lower atmosphere the turbulent diffusion coefficient is usually larger than $10^5$  cm$^2$  s$^{-1}$. Therefore, for aerosols smaller than 2.5 $\mu$m the turbulent thermal diffusion ratio is   $\alpha_{_{TD}} \approx 1$ (as well as for gases), while for coarser aerosols the TTD ratio $\alpha_{_{TD}}$  varies from 5 to 10 (see Figures 6 and 7). Therefore, the same temperature gradient affects the dispersion of coarse aerosols 5-10 times stronger than that of the fine particles. This is also evident from a simple fact that both velocities, $V^{\rm eff}_z$ and the sedimentation $W_g$, are proportional to the squared particle size.

\section{Discussion}

The phenomenon of turbulent thermal diffusion has been predicted theoretically for the atmospheric and laboratory turbulent flows and then detected in a series of laboratory experiments. Therefore, there is no doubt in the existence and importance of this effect in case of strong temperature gradients. The goal of this study is to assess the importance of this phenomenon in the atmosphere and elucidate where one can find the atmospheric signature of this effect. The main problem is that, contrary to the controlled laboratory experiments, the influence of TTD on aerosols distribution in the Earth atmosphere has to be distinguished from other phenomena affecting aerosol profiles.

The most important phenomena affecting the aerosol profile near the tropopause, are as follows:

(i) aerosol dynamics;

(ii) cloud microphysics;

(iii) dynamic interaction of the vertical gradients of the turbulent diffusion coefficient $K_D$ and aerosol concentrations;

(iv) gravitational sedimentation of particles;

(v) various dynamic phenomena violating the steady-state assumption and disturbing the aerosol profiles, e.g., tropical deep convection;

(vi) variation of the aerosol concentration profiles caused by the Bernoulli effect and the wind speed gradient near the tropopause.

The contributions of these effects to the aerosol profile strongly vary and have to be evaluated case-by-case. Quantitative evaluation of impact of these effects often requires such details about aerosol composition and size spectrum, as well as about the atmospheric conditions, that are not available. However, some conclusions can be drawn from the presented data.

There are several factors which indicate that the mechanism related to TTD is universal and widespread, and should exist at least in low- and mid- latitudes. They include:

(i) striking similarity of the aerosol profiles in tropics and mid-latitudes;

(ii) very large fraction of considered aerosol profiles reveal the temperature-concentration anti-correlation and faithfully follow the functional dependence given by Eq.~(\ref{BB18}). The theory of TTD provides the explanation for this anti-correlation and even predicts this functional dependence.

In the subsequent qualitative analysis, we explore the possible alternative mechanisms which can maintain such aerosol profiles stable over a long time period. It must be emphasized that we do not consider the mechanisms of formation and transport of these amounts of particles to high altitudes. These can be associated with numerous effects, which are much faster and stronger then TTD. Our analysis is focussed on the mechanism which can explain the existence the observed long-living aerosol concentrations in different parts of the globe.

Formation of the cirrus clouds can be of some importance for tropical regions where the situation is quite ambiguous but it is hardly of importance for the mid-latitude GOMOS profiles. Notably, the aerosol optical characteristics are closer to sulphates than to water ice crystals [see Vanhellemont et al., 2005]. Therefore, cirrus clouds formation cannot be the main mechanism responsible for the observed aerosol profiles. It should be also kept in mind that the GOMOS observations are feasible only in the absence of thick clouds which, consequently, cannot affect the aerosol concentration profiles.

The dynamic interaction of aerosol concentration near the tropopause with the fast-changing $K_D$ with altitude follows from expression: $\nabla_z (K_D \nabla_z N) = (\nabla_z K_D) (\nabla_z N) + K_D \nabla_z^2 N$. The first term in the r.h.s. of this equation corresponds to advection of the species in the direction opposite to the gradient of the diffusion coefficient. This would lead to temporal accumulation of the particles near the altitude with sharp changes of the turbulent diffusion coefficient $K_D$, which is closely related to local temperature gradients. However, this explanation also has several limitations.

Firstly, the temperature minimum at the tropopause exists only for the absolute temperature while the potential temperature is steadily growing with altitude in the upper troposphere and the lower stratosphere. Consequently, the stratification above and below the minimum of the temperature is stable, which implies $K_D$ is comparatively small in the considered height range and its gradient is also not large.

Secondly, Eqs.~(\ref{BB4}) and~(\ref{BB5}) in their "classical" form with $V^{\rm eff}_z=0$  do not have a steady-state solution with a maximum of concentration inside a modelled domain. The increased concentration at the top of turbulent domain can occur only in case of strong elevated aerosol sources and will not remain for a long time. Eventually, this maximum will be smeared by turbulence and finally will disappear. Sedimentation velocity, that is substantial at these altitudes due to low air density and viscosity, would speed-up this process. However, numerous randomly chosen GOMOS observations (in addition to the ones reported above) show that the aerosol layers in the vicinity of the temperature minimum are commonly occurring.

In general, the strong correlation between the temperature and aerosol concentration profiles cannot be explained by any of the dynamic processes whose main driving forces are associated with the wind and turbulence rather than temperature per-se. Therefore, such mechanisms cannot cause the observed functional dependence between temperature and aerosol concentrations. Hence, the effects, such as Bernoulli jet, are also hardly of primary importance here.

Formation of new aerosols and deep convection in tropical regions are probably the most-prominent of the remaining alternative mechanisms of the aerosol appearance at high altitudes. However, neither of them explains the aerosol persistence and formation of the profiles anti-correlated with temperature. Indeed, sedimentation of particles is substantial at these altitudes, and such downward flux of particles would be readily observable and would result in essentially different forms of the relation~(\ref{BB18}) for the parts of the aerosol profiles above and below the temperature minimum. Since nothing of this kind is observed, we have to assume that some other mechanism is responsible for formation of the vertical particle profile after aerosols have been formed or transported by, e.g., deep convection towards the tropopause.

The above qualitative analysis shows that none of the considered effects can explain the observed stable in time widespread functional relation between the temperature and aerosol concentrations given by Eq.~(\ref{BB18}) with $\alpha_{_{TD}}$  given by Eq.~(\ref{BB16}) (see numerous examples in Figures~6-8). Validity of these relations for the analyzed hundreds of cases is probably the main argument in favour of the phenomenon of turbulent thermal diffusion as one of the main mechanisms that stabilizes the observed anti-correlation between the temperature and aerosol concentration profiles, and renders this anti-correlation a widespread feature of aerosol profiles at low and middle latitudes.

It is worth mentioning that the near-tropopause conditions are drastically different from the laboratory experiments [see Eidelman et al., 2004; 2006a, 2006b; Buchholtz et al., 2004]. At normal pressure, the typical values of $\alpha_{_{TD}}$  measured in the laboratory experiments do not exceed 3-4, while the near-tropopause observations showed much higher values. The main reason for this difference is substantially lower air density and viscosity, as well as the smaller turbulence intensity. With corresponding corrections taken into account, the theoretical predictions for $\alpha_{_{TD}}$ appear to be in a fairly good agreement with the observations (see Figures 6-8).

Detection of the turbulent thermal diffusion in the observations in the atmospheric boundary layer is even more complicated than in the middle atmosphere. Our modelling experiments showed that characteristic times of the competing effects in the boundary layer are significantly shorter than those required for approaching the steady-state solution with the evident signature of the turbulent thermal diffusion. In addition, the TTD ratio $\alpha_{_{TD}}$  is substantially lower near the surface than in the upper troposphere, thus requiring stronger temperature gradients for the effect to be easily detectable. Consequently, a comprehensive analysis of profile-resolving observations is required for a detection and quantification of the phenomenon of turbulent thermal diffusion in the lower troposphere.

The numerical experiments with the SILAM model provided the first estimates for the TTD impact in the lower troposphere and highlighted its main features. Moderate but systematic changes in the aerosol distributions that emerged from the simulations seem to comply well with the theory of this phenomenon. Indeed, TTD affects the whole mass of aerosols usually acting opposite to gravitational sedimentation and being comparable with it. For some naturally occurring temperature gradients and intensity of turbulence, sedimentation of fine particles can be fully compensated by  $V^{\rm eff}_z$, essentially canceling their dry deposition (the diffusion-driven dry deposition is usually insignificant for coarse particles). In our numerical simulations, the average effect of the turbulent thermal diffusion was at the level of 10-15 \% with general tendency of redistribution of aerosol masses upwards. As an immediate consequence, the aerosol transport distance has increased as well as the concentrations over the elevated and remote regions (see Figure~10).

As can be seen from the maps shown in Figure~10, the relative effect of TTD on coarse particles is considerably stronger than that on fine aerosols. Indeed, the TTD term [see Eqs.~(\ref{BB5}), (\ref{BB9}) and~(\ref{BB11})] has a contribution which is proportional to the squared particle size. Consequently, for particles with small sedimentation velocity $W_g$ the absolute impact of TTD is also small and easily overshadowed by other effects, such as macro-scale turbulence, transport with mean vertical wind, etc. For coarser particles, the sedimentation is much more important, therefore its adjustment due to TTD is also more pronounced.

Notably, Eq.~(\ref{BB9}) for the effective velocity, $V^{\rm eff}_z$, for for non-inertial particles or gaseous admixtures (i.e., for $\alpha_{_{TD}}=1)$ has been already applied by Atreya et al. [1999] for study of vertical mixing in the atmospheres of Jupiter and Saturn. This equation was obtained by Atreya et al. [1999] using phenomenological arguments.

The theory of the phenomenon of turbulent thermal diffusion still needs to be refined by more tests in carefully controlled experiments, atmospheric observations and in direct numerical simulations of density-driven turbulent flow, where the flow physics can be controlled and modified to test the different aspects of the theory.

\section{Conclusions}

This study is a follow-up of theoretical investigations and laboratory experiments [Elperin et al., 1996; 1997a; Eidelman et al., 2004, 2006a, 2006b], which demonstrated the existence of phenomenon of turbulent thermal diffusion. This paper provides the first observational evidence and quantitative evaluation of the importance of TTD in the atmosphere.

The analytical part of this paper provides a set of practically applicable formulations, directly useable for analysis of observations and ready for implementation in dispersion models. The most important effect is the extra term in the diffusion equation expressed through macro-scale parameters of turbulence and aerosol features, see Eqs.~(\ref{BB5}), (\ref{BB9})-(\ref{BB11}) and~(\ref{BB16}).

Application of these formulations to the analysis of GOMOS observations near the tropopause seems to explain otherwise confusing shape of aerosol vertical profiles with elevated concentrations located almost symmetrically to the temperature profile. To the best knowledge of the authors, such symmetry has not been explained before.

It must be emphasized that the characteristic times and the magnitude of turbulent thermal diffusion are insufficient for formation of aerosol layers. Numerical simulations have confirmed that the impact of TTD is systematic but moderate, which also complicates detection of this phenomenon among dynamic processes in the boundary layer. Hence, turbulent thermal diffusion alone is insufficient for elevating substantial amount of aerosols to the tropopause and formation of the aerosol layers with the enhanced concentration of particles. In the performed modelling experiment, the simulation time and/or the size of the computational domain were insufficient for the formation of aerosol concentrations. Consequently, it is plausible to expect that turbulent thermal diffusion hardly plays the main role in the formation of the near-tropopause aerosol layers. Other effects, such as deep convection, buoyant plumes from wild-land fires, formation of new aerosols, etc., may be more important in this respect. However, turbulent thermal diffusion is, probably, the only effect that can trap the aerosols in the vicinity of the minimum of temperature once they are formed.

\appendix

\section{The effective velocity for small Peclet numbers}

In this Appendix we derive Eqs.~(\ref{BB5}) and~(\ref{BB6b})  using a quasi-linear approach or a second order correlation approximation [see, e.g., Moffatt, 1978]. We rewrite Eq.~(\ref{BB7}) in the following form:
\begin{eqnarray}
{\partial n \over \partial t} + \bec\nabla {\bf \cdot} \, {\bf Q} - D \bec{\nabla}^2 n  = I \;,
\label{C1}
\end{eqnarray}
where ${\bf Q} = {\bf v} n - \langle {\bf v} n \rangle$  is the nonlinear term and $I = N \, \bec\nabla {\bf \cdot} \, {\bf v} +  ({\bf v} {\bf \cdot} \bec{\nabla}) N$ is the source term. Let us neglect the nonlinear term but keep the molecular diffusion term in Eq.~(\ref{C1}).
For this reason this approach is called a quasi-linear or perturbational approach. This approximation for a given velocity field is valid only for small Peclet numbers (${\rm Pe} \ll 1$), where ${\rm Pe} = u_{0} \, \ell / D$. Let us rewrite Eq.~(\ref{C1}) in a Fourier space. Then the solution of Eq.~(\ref{C1}) is given by $n(\omega, {\bf k}) = G_D(\omega, {\bf k}) I(\omega, {\bf k})$, where $G_D(\omega, {\bf k}) = (D k^2 + i \omega)^{-1}$.

We apply a standard two-scale approach, i.e., the non-instantaneous two-point second-order correlation function can be written as follows:
\begin{eqnarray}
&& \langle v_i(t_1, {\bf x}) \, n (t_2, {\bf  y}) \rangle = \int \langle v_i (\omega_1, {\bf k}_1) n (\omega_2, {\bf k}_2) \rangle \exp[i({\bf  k}_1 {\bf \cdot} {\bf x}
\nonumber\\
&& \quad \quad + {\bf k}_2 {\bf \cdot} {\bf y}) + i(\omega_1 t_1 + \omega_2 t_2)] \,d\omega_1 \, d\omega_2 \,d{\bf k}_1 \, d{\bf k}_2
\nonumber\\
&& \quad \quad = \int F_i^{(n)}(\omega, {\bf k})  \exp[i {\bf k} {\bf \cdot} {\bf r} + i\omega \, \tilde \tau] \,d\omega \,d {\bf k} \;,
\label{C2}
\end{eqnarray}
where we use large scale variables: ${\bf R} = ({\bf x} + {\bf y}) / 2$, $\, {\bf K} = {\bf k}_1 + {\bf k}_2$, $\, t = (t_1 + t_2) / 2$, $\, \Omega = \omega_1 + \omega_2$, and small scale  variables: ${\bf r} = {\bf x} - {\bf y}$, $\, {\bf k} = ({\bf k}_1 - {\bf k}_2) / 2$, $\, \tilde \tau = t_1 - t_2$, $\, \omega = (\omega_1 - \omega_2) / 2$ [see, e.g., Roberts and Soward, 1975]. Here
\begin{eqnarray}
&& F_i^{(n)}(\omega, {\bf k}) = \int \langle v_i(\omega_1, {\bf k}_1) \, n(\omega_2, {\bf k}_2) \rangle \exp[i \Omega t
\nonumber\\
&& \quad \quad\quad\quad + i {\bf K} {\bf \cdot} {\bf R}] \,d \Omega \,d {\bf  K} \;,
\label{C3}
\end{eqnarray}
and $\omega_1 = \omega + \Omega / 2$, $\, \omega_2 = - \omega + \Omega / 2$, ${\bf k}_1 = {\bf k} + {\bf  K} / 2$, $\, {\bf k}_2 = - {\bf k} + {\bf  K} / 2$. We assume here that there exists a separation of scales,
i.e., the maximum scale of random motions $\ell$ is much
smaller than the characteristic scales of inhomogeneities of the
mean particle number density and mean fluid density.

Hereafter we use the simplest model for the second moment of a random velocity field in a Fourier space:
\begin{eqnarray}
&& \langle u_i(\omega, {\bf k}) \, u_j(-\omega, -{\bf k}) \rangle = {u_0^2 \, E(k) \over 8 \pi k^2} \Big[\delta_{ij} - {k_i \, k_j \over k^2}
\nonumber\\
&& \quad \quad\quad\quad\quad\quad + {i \over k^2} \, \big(\Lambda_i k_j - \Lambda_j k_i\big)\Big] \, \delta(\omega) \;,
\label{C4}
\end{eqnarray}
that satisfies the continuity equation in anelastic approximation $\bec\nabla {\bf \cdot} \, {\bf u} = u_i \, \Lambda_i$, where $\Lambda_i = - \nabla_i \rho / \rho$, $\, \delta(\omega)$ is the Dirac's delta-function, $\, \delta_{ij}$ is the Kronecker tensor, the energy spectrum function is $E(k) = k_0^{-1} \, (q-1) \, (k / k_{0})^{-q}$, the exponent $1<q<3$, the wave number $k_{0} = 1 / \ell$, the length $\ell$ is the maximum scale of random motions and $u_0$ is the characteristic velocity in the maximum scale of random motions. Hereafter we neglect the small terms $\sim O[(\Lambda \ell)^2]$.

We consider non-inertial particles, i.e., ${\bf v} = {\bf u}$. Then the turbulent flux of particles is given by
\begin{eqnarray}
F_i^{(n)} = \int \langle u_i(\omega, {\bf k}) \, I(- \omega, - {\bf k}) \rangle \, G_D(- \omega, - {\bf k}) \,d \omega \,d {\bf  k} .
\label{C5}
\end{eqnarray}
After integration in $\omega$-space and in ${\bf k}$-space we arrive at Eq.~(\ref{BB5}) for the turbulent flux of particles with
\begin{eqnarray}
K_D &=& {(q-1) \, u_{0} \, \ell \over 3(q+1)} \, {\rm Pe} \;,
\label{C6}\\
V^{\rm eff}_i &=& K_D {\nabla_i \rho  \over \rho} = - K_D {\nabla_i T  \over T} \; .
\label{C7}
\end{eqnarray}
In the derivation we used the integral:
\begin{eqnarray*}
\int_0^\pi \sin \, \theta \, d\theta \int_0^{2\pi} \, (k_i \, k_j / k^2) \, d\varphi = (4\pi/3) \, \delta_{ij} \; .
\end{eqnarray*}

\section{The effective velocity for large Peclet numbers}

In this Appendix we derive Eqs.~(\ref{BB5}) and~(\ref{BB6b})  using the $\tau$ approach that is valid for large Peclet and Reynolds numbers. Using Eq.~(\ref{C1}) written in a Fourier space we derive equation for the instantaneous two-point second-order correlation function $F_i^{(n)}(t, {\bf k}) = \langle u_i(t, {\bf k}) \, n(t, -{\bf k}) \rangle$:
\begin{eqnarray}
{dF_i^{(n)} \over dt} = \langle u_i(t, {\bf k}) \, I(t, -{\bf k}) \rangle + \hat{\cal M} F_i^{(III)}({\bf k}) \;,
\label{D1}
\end{eqnarray}
where $\hat{\cal M} F_i^{(III)}({\bf k}) = - [\langle u_i \, \bec\nabla {\bf \cdot} \, {\bf Q} \rangle + \langle (\partial u_i / \partial t) \, n \rangle - D \langle u_i \, \bec{\nabla}^2 n \rangle]_{\bf k}$ are the third-order moment terms appearing due to the nonlinear terms which include also molecular diffusion term.

The equation for the second moment includes the first-order spatial
differential operators $\hat{\cal M}$  applied to the third-order
moments $F^{(III)}$. A problem arises how to close the system, i.e.,
how to express the third-order terms $\hat{\cal M}
F^{(III)}$ through the lower moments $F^{(II)}$ [see, e.g.,
Orszag, 1970; Monin and Yaglom, 1975; McComb 1990]. We use the spectral $\tau$ approximation which postulates that the deviations of the third-moment terms, $\hat{\cal M} F^{(III)}({\bf k})$, from the contributions to these terms afforded by the background turbulence, $\hat{\cal M} F^{(III,0)}({\bf k})$, can be expressed through the similar deviations of the second moments, $F^{(II)}({\bf k}) - F^{(II,0)}({\bf k})$:
\begin{eqnarray}
&& \hat{\cal M} F^{(III)}({\bf k}) - \hat{\cal M} F^{(III,0)}({\bf
k})
\nonumber\\
&& \quad \quad\quad = - {1 \over \tau_r(k)} \, \Big[F^{(II)}({\bf k}) - F^{(II,0)}({\bf k})\Big] \;,
\label{D2}
\end{eqnarray}
[see, e.g., Orszag, 1970; Pouquet et al., 1976; Elperin et al., 2006], where $\tau_r(k)$ is the scale-dependent relaxation time, which can be identified with the correlation time $\tau(k)$ of the turbulent velocity field for large Reynolds and Peclet numbers. The functions with the superscript $(0)$ correspond to the background turbulence with a zero gradient of fluid density and a zero gradient of the mean number density of particles. Validation of the $\tau$ approximation for different situations has been performed in numerous numerical simulations and analytical studies [see, e.g., review by Brandenburg and Subramanian, 2005; and also discussion in Rogachevskii and Kleeorin, 2007, Sec. 6].

Note that the contributions of the terms with the superscript $(0)$ vanish because when the gradients of fluid density and the gradients of the mean number density are zero, the flux of particles vanishes. This implies that Eq.~(\ref{D2}) reduces to $\hat{\cal M} F_i^{(III)}({\bf k}) = - F_i^{(n)}({\bf k}) / \tau(k)$. We also assume that the characteristic time of variation of the second moment $F_i^{(n)}({\bf k})$ is substantially larger than the correlation time $\tau(k)$ for all turbulence scales. Therefore, in a steady-state Eq.~(\ref{D1}) yields the turbulent flux of particles $F_i^{(n)} = \int \tau(k) \langle u_i(t, {\bf k}) \, I(t, -{\bf k}) \rangle \, d{\bf k}$. Now we use the following simple model for the second moment of turbulent velocity field:
\begin{eqnarray}
&& \langle u_i({\bf k}) \, u_j(-{\bf k}) \rangle = {u_0^2 \, E(k) \over 8 \pi k^2} \Big[\delta_{ij} - {k_i \, k_j \over k^2}
\nonumber\\
&& \quad \quad\quad\quad\quad\quad + {i \over k^2} \, \big(\Lambda_i k_j - \Lambda_j k_i\big)\Big] \; .
\label{D3}
\end{eqnarray}
After integration in ${\bf k}$-space we arrive at Eq.~(\ref{BB5}) for the turbulent flux of particles with
\begin{eqnarray}
K_D &=& {u_{0} \, \ell \over 3} \;,
\label{D4}\\
V^{\rm eff}_i &=& K_D {\nabla_i \rho  \over \rho} = - K_D {\nabla_i T  \over T} \; .
\label{D5}
\end{eqnarray}
In the derivation we used the following expression for the turbulent correlation time, $\tau(k) = 2 \, t_T \, (k / k_{0})^{1-q}$, where $t_T = \ell / u_{0}$ is the characteristic turbulent time. Comparison of Eqs.~(\ref{C7}) and ~(\ref{D5}) shows that the form of the effective velocity $V^{\rm eff}_i$ is the same for small and large Peclet numbers, while the values of turbulent diffusion coefficients are different in these two cases.

\section{The budget equation for the turbulent flux of particles}

Let us consider inertial particles suspended in the turbulent fluid. Their concentration is characterized by the mean value, $N$, and fluctuation, $n$, of the number density of particles [m$^{-3}$]. The budget equation for the turbulent flux of particles,  $F_i^{(n)} = \langle v_i \, n \rangle$, reads
\begin{eqnarray}
{DF_i^{(n)} \over Dt} + \nabla_j \Phi_{ij}^{(n)} &=& -\langle u_i \, u_j \rangle \, \nabla_j N - F_j^{(n)} \nabla_j U_i + Q_i^{(n)}
\nonumber\\
&&- \langle u_i \, \bec\nabla {\bf \cdot} \, {\bf v} \rangle N - \varepsilon_i^{(n)} \;,
\label{A1}
\end{eqnarray}
where $D / Dt = \partial / \partial t + {\bf U} {\bf \cdot} \bec{\nabla} $, $\; {\bf v}$  is the particle velocity, ${\bf u}$ and ${\bf U}$ are the fluctuations and mean components of the fluid velocity, $\Phi_{ij}^{(n)}$  is the third-order turbulent flux of $F_i^{(n)}$:
\begin{eqnarray}
\Phi_{ij}^{(n)} = \langle u_i \, u_i \, n\rangle + {1 \over \rho} \, \langle p \, n \rangle \, \delta_{ij} \;,
\label{A2}
\end{eqnarray}
$\varepsilon_i^{(n)}$ is the molecular dissipation rate of $F_i^{(n)}$:
\begin{eqnarray}
\varepsilon_i^{(n)} = - D \, \langle u_i \, \bec{\nabla}^2 n\rangle - \nu \, \langle n \, \bec{\nabla}^2 u_i  \rangle \;,
\label{A3}
\end{eqnarray}
$\nu$ is the kinematic viscosity of fluid, $D$ is the Brownian diffusivity of particles and  $p$ are the fluid pressure fluctuations. The Kolmogorov closure hypothesis implies that
\begin{eqnarray}
\varepsilon_i^{(n)} = {F_i^{(n)} \over C_n t_T} = {F_i^{(n)} \, E_z^{1/2} \over C_n \, \ell_z} \;,
\label{A4}
\end{eqnarray}
where $\ell_z$  is the vertical turbulent length scale, $E_z$  is the vertical turbulent kinetic energy, $t_T = \ell_z / E_z^{1/2}$  is the turbulent time,  $C_n$ is an empirical dimensionless coefficient. The term $Q_i^{(n)}$ in Eq.~(\ref{A1}) is given by the following expression:
\begin{eqnarray}
Q_i^{(n)} = {1 \over \rho} \, \langle p \, \nabla_i n \rangle + \beta e_i \, \langle n \, \theta_p \rangle \;,
\label{A5a}
\end{eqnarray}
$\theta_p$  is the potential temperature fluctuation, ${\bf e}$  is the vertical unit vector, $\beta=g/T_\ast$  is the buoyancy parameter, ${\bf g}$  is the acceleration of gravity, and  $T_\ast$ is a reference value of the mean temperature $T$  and $\rho$ is the fluid density. In Appendix D we derive the following expression for the correlation term $Q_i^{(n)}$:
\begin{eqnarray}
Q_i^{(n)} &=& - C_D \, e_i \, {\beta \ell_z \over E_z^{1/2}} \, \big(F_j \, \nabla_j N + \langle \theta_p \, \bec\nabla {\bf \cdot} \, {\bf v} \rangle \, N
\nonumber\\
&& + F_j^{(n)} \, \nabla_j \Theta \big) \;,
\label{A5b}
\end{eqnarray}
where $C_D$  is an empirical dimensionless constant, $F_i = \langle u_i \, \theta_p\rangle$  is the flux of the potential temperature and $\Theta$  is the mean potential temperature (or the mean virtual potential temperature accounting for specific humidity), that is defined as $\Theta = T (P_\ast / P)^{1-\gamma^{-1}}$. Here $P$  is the mean pressure,  $P_\ast$ is its reference value, and $\gamma = c_p/c_v=1.41$ is the specific heat ratio. Then Eq.~(\ref{A1}) can be rewritten
\begin{eqnarray}
{DF_i^{(n)} \over Dt} &+&  \nabla_j \Phi_{ij}^{(n)} = -\tau_{ij} \nabla_j N - F_j^{(n)} \nabla_j U_i
\nonumber\\
&-& \langle u_i \bec\nabla {\bf \cdot} \, {\bf v} \rangle N - {F_i^{(n)} \, E_z^{1/2} \over C_n \, \ell_z} - C_D \,  {\beta \ell_z \over 2 E_z^{1/2}} \, \nonumber\\
&\times& \big[ (e_i \, F_j + e_j \, F_i) \, \nabla_j N + 2 e_i \, \langle \theta_p \, \bec\nabla {\bf \cdot} \, {\bf v} \rangle \, N
\nonumber\\
&+& (e_i \,F_j^{(n)} + e_j \,F_i^{(n)}) \, \nabla_j \Theta \big] \;,
\label{A6}
\end{eqnarray}
where the r.h.s. describes the local budget of the turbulent flux of particles (aerosols),  $F_i^{(n)}$.
The third-order flux $\Phi_{zz}^{(n)}$  can be expressed using the turbulent diffusion approximation:
\begin{eqnarray}
\Phi_{zz}^{(n)} &=& - C_{F} \, K \, \nabla_z F_z^{(n)} \;,
\label{A7}
\end{eqnarray}
where $K= \ell_z \, E_z^{1/2}$. Then the budget equation for the vertical particle flux, $F_z^{(n)}$, reads:
\begin{eqnarray}
{DF_z^{(n)} \over Dt} &+& C_{F} \,  \nabla_z (K \, \nabla_z F_z^{(n)}) = - {N \over \rho} \, \tau_s \, \langle u_z \, \bec{\nabla}^2 p \rangle
\nonumber\\
&-& \Big(2 E_z - C_D \,  {\beta \, F_z \, \ell_z \over E_z^{1/2}} \Big)
\Big(\nabla_z N - N {\nabla_z \rho \over \rho} \Big)
\nonumber\\
&-& \Big(C_D \,  \beta \, \nabla_z \Theta + {E_z \over C_n \, \ell_z^2} \Big) \, {\ell_z \, F_z^{(n)} \over E_z^{1/2}} \;,
\label{A8}
\end{eqnarray}
where $\tau_s$ is the particle Stokes time and $C_F$  is the empirical dimensionless constant. Equation~(\ref{A8}) is a complementary equation to the non-local closure model suggested by Zilitinkevich et al. [2007]. In the steady-state, homogeneous regime of turbulence, Eq.~(\ref{A8}) yields the vertical component of the turbulent flux of particles in the form of Eq.~(\ref{BB5}) with the coefficients $K_D$ and $K_{TD}$ given by Eqs.~(\ref{BB10}) and~(\ref{BB11}). Equation~(\ref{A6}) also allows us to determine the horizontal components of the particle flux [see Eq.~(\ref{A9})].

Now let us derive Eq.~(\ref{I2}). The effective velocity ${\bf V}^{\rm eff} $ depends on the turbulent heat flux $\langle {\bf u} \, \theta \rangle$ that is determined by the well known equation: $\langle {\bf u} \, \theta \rangle  = - K_H \, \bec{\nabla} T$ [see, e.g., Monin and Yaglom, 1975], where $K_H \sim u_0 \, \ell / 3 $ is the coefficient of turbulent thermal conductivity. Note that herein we do not consider situation with very high gradients when gradient transport assumption is violated. The above formula for the mean turbulent heat flux is written in the ${\bf r}$-space. The corresponding second moment in ${\bf k}$-space is given by $\langle u_i ({\bf k}) \, \theta (-{\bf k}) \rangle = - \tau(k) \, \langle u_i({\bf k}) \, u_j(-{\bf k}) \rangle \nabla_j T$, where the second moment of turbulent velocity field is given by Eq.~(\ref{D3}), the turbulent correlation time is $\tau(k) = 2 \, t_T \, (k / k_{0})^{-2/3}$, the energy spectrum function is $E(k) = (2/3) k_0^{-1} \, (k / k_{0})^{-5/3}$.

Multiplying equation for $\langle u_i ({\bf k}) \, \theta(-{\bf k}) \rangle $ by $- k^2 \tau(k)$ and integrating in ${\bf k}$-space we obtain $t_T \, \langle {\bf u} \, \bec{\nabla}^2 \theta \rangle = (2/3) \, \ln({\rm Re}) \, B({\rm Re}, a_{\ast}) \,\bec{\nabla} T$, where the function $B({\rm Re}, a_{\ast}) = 1$ when the particle size  $a_{\ast} < a_{\rm cr}$, and $B({\rm Re}, a_{\ast}) = 1 - 3 \ln(a_{\ast} / a_{\rm cr}) / \ln({\rm Re})$ when $a_{\ast} \geq a_{\rm cr}$ [see Elperin et al., 2000a]. Here $a_{\rm cr}$ is the critical particle size that is given by $a_{\rm cr} = \ell_{\nu} (\rho / \rho_{p})^{1/2}$ and $\ell_{\nu} = \ell \, {\rm Re}^{-3/4} $ is the Kolmogorov viscous scale of turbulence.

\section{Derivation of the budget equation for $\langle n \, \theta_p \rangle$ and Eq.~(\ref{A5b})}

Equations for fluctuations of the potential temperature and the particle number density read:
\begin{eqnarray}
{D\theta_p \over Dt} &=& - u_{j} \nabla_j \, (\Theta + \theta_p) + \kappa \, \bec{\nabla}^2 \theta_p \;,
\label{B1}\\
{Dn \over Dt} &=& - (u_{j} \nabla_j + \bec\nabla {\bf \cdot} \, {\bf v}) \, (N + n) - n \, \bec\nabla {\bf \cdot} \, {\bf V} + D \, \bec{\nabla}^2 n \;,
\nonumber\\
\label{B2}
\end{eqnarray}
where  ${\bf V}$ is the mean particle velocity. Multiplying Eq.~(\ref{B1}) by $n$ and Eq.~(\ref{B2}) by $\theta_p$, averaging and adding yield the budget equation for the correlation function $E_{n\theta} = \langle n \, \theta_p \rangle$:
\begin{eqnarray}
{D E_{n\theta} \over Dt} &+& \nabla_j \Phi_{j}^{(n\theta)} = - F_j^{(n)} \nabla_j \Theta - F_j \nabla_j N
\nonumber\\
&-& \langle \theta_p \, \bec\nabla {\bf \cdot} \, {\bf v} \rangle N - E_{n\theta} \, \bec\nabla {\bf \cdot} \, {\bf V} - \varepsilon^{(n\theta)} \;,
\label{B3}
\end{eqnarray}
Here $\Phi_{j}^{(n\theta)}$  is the third-order turbulent flux:
\begin{eqnarray}
\Phi_{i}^{(n\theta)} = \langle u_i \, n \, \theta_p \rangle \;,
\label{B4}
\end{eqnarray}
and $\varepsilon^{(n\theta)}$ is the molecular dissipation rate of $E_{n\theta}$:
\begin{eqnarray}
\varepsilon^{(n\theta)} = - D \, \langle \theta_p \, \bec{\nabla}^2 n\rangle - \kappa \, \langle n \, \bec{\nabla}^2 \theta_p  \rangle \; .
\label{B5}
\end{eqnarray}
Molecular dissipation rate $\varepsilon^{(n\theta)}$ can be expressed using the Kolmogorov closure hypothesis:
\begin{eqnarray}
\varepsilon^{(n\theta)} = {E_{n\theta} \over C_{n\theta} t_T} = {E_{n\theta} \, E_z^{1/2} \over C_{n\theta} \, \ell_z} \;,
\label{B6}
\end{eqnarray}
where $C_{n\theta}$ is the dimensionless constant.

In the steady-state, homogeneous regime of turbulence, Eq.~(\ref{B3}) reduces to the turbulent diffusion formulation:
\begin{eqnarray}
E_{n\theta} &=& - C_{n\theta} \,  {\ell_z \over E_z^{1/2}} \, \big(F_j^{(n)} \nabla_j \Theta + F_j \nabla_j N
\nonumber\\
&& + \langle \theta_p \, \bec\nabla {\bf \cdot} \, {\bf v} \rangle N \big) \;,
\label{B7}
\end{eqnarray}
where we consider only gradient approximation neglecting higher spatial derivatives.

Now let us determine the correlation term $Q_i^{(n)}$  which includes the correlation function $\rho^{-1} \, \langle p \, \nabla_i n \rangle$ (see Eq.~(\ref{A5a})). To this end we use the following identities:
\begin{eqnarray}
\rho^{-1} \, p &=& \beta \, \Delta^{-1} \nabla_z \theta_p \;,
\nonumber\\
\rho^{-1} \, \langle \theta_p \, \nabla_z p \rangle &=& - \beta \, \langle \theta_p \, \Delta^{-1} \nabla_z^2 \theta_p \rangle \;,
\label{B8}
\end{eqnarray}
which follow from the Navier-Stokes  equation. Indeed, calculating the divergence of the Navier-Stokes yields:
\begin{eqnarray}
\rho^{-1} \, \bec{\nabla}^2 p = - \beta \, \nabla_z \theta_p \; .
\label{B9}
\end{eqnarray}
Applying the inverse Laplacian to Eq.~(\ref{B9}) we arrive at Eq.~(\ref{B8}). Therefore,
\begin{eqnarray}
\rho^{-1} \, \langle p \, \nabla_i n \rangle &=& \beta \, \langle (\nabla_i n) \, \Delta^{-1} \nabla_i \theta_p \rangle =
\beta \, \nabla_i \, \langle n \, \Delta^{-1} \nabla_z \theta_p \rangle \nonumber\\
&& - \beta \, \langle n \, \Delta^{-1} \nabla_z \nabla_i \theta_p \rangle
\;,
\label{B10}
\end{eqnarray}
Our analysis shows that the last term in Eq.~(\ref{B10}) can be estimated as:
\begin{eqnarray}
\langle n \, \Delta^{-1} \nabla_z^2 \theta_p \rangle
\approx E_{n\theta} \, (1 + \delta^{-1}) \, \Big[1 - {\arctan \sqrt{\delta} \over \sqrt{\delta}} \, \Big] \;,
\nonumber\\
\label{B11a}
\end{eqnarray}
[see  Zilitenkevich et al., 2007], where $\delta = l_h^2 / l_z^2$  and  $l_z$ and $l_h$ are the correlation lengths of the correlation function $\langle n(t, {\bf x}) \, \theta_p(t, {\bf y}) \rangle$  in the vertical and horizontal directions. For nearly isotropic case $(\delta \ll 1)$, Eq.~(\ref{B11a}) reads:
\begin{eqnarray}
\langle n \, \Delta^{-1} \nabla_z^2 \theta_p \rangle
\approx {1 \over 3} \, \Big(1 + {2\delta \over 5}\Big) \, E_{n\theta} \;,
\label{B11b}
\end{eqnarray}
while for a strongly anisotropic case $(\delta \gg 1)$, Eq.~(\ref{B11a}) yields:
\begin{eqnarray}
\langle n \, \Delta^{-1} \nabla_z^2 \theta_p \rangle
\approx \Big(1 - {\pi \over 2\sqrt{\delta}} + {2 \over \delta}\Big) \, E_{n\theta} \; .
\label{B11c}
\end{eqnarray}
Equations~(\ref{B7})-(\ref{B11c}) yield the expression for the correlation term $Q_i^{(n)} = \rho^{-1} \, \langle p \, \nabla_i n \rangle + \beta e_i \, \langle n \, \theta_p \rangle$ given by Eq.~(\ref{A5b}).

\begin{acknowledgments}
This work has been supported by Israel Science Foundation governed by the Israeli Academy of Science (Grant 259/07), the EC FP7 projects ERC PBL-PMES (No. 227915) and MEGAPOLI (No. 212520), EU Project FUMAPEX EVK4-CT-2002-00097, ARO Project W911NF-05-1-0055, EU-GEMS, TEKES-KOPRA and the Academy of Finland projects (POLLEN, IS4FIRES, postdoctoral researcher project of VS). The authors thank ESA and the GOMOS team for the GOMOS data.
\end{acknowledgments}

\end{document}